\documentclass[pra,twocolumn,nofootinbib]{revtex4}
\usepackage{graphicx}
\usepackage{hyperref}
\usepackage[section]{placeins}
\usepackage{amsmath}
\usepackage{amssymb}
\usepackage{dsfont}
\usepackage{bbold}
\usepackage{color}
\newcommand{\ket}[1]{|#1\rangle}             

\begin{document}

\title{Inferring causal structure: a quantum advantage}

\author{Katja Ried$^{1,2,3\ast}$, Megan Agnew$^{1,2\ast}$, Lydia Vermeyden$^{1,2}$, Dominik Janzing$^4$, Robert W. Spekkens$^{3}$ and Kevin J. Resch$^{1,2}$}
\affiliation{$^1$Institute for Quantum Computing, University of Waterloo, Waterloo, Ontario, Canada, N2L 3G1}
\affiliation{$^2$Department of Physics \& Astronomy, University of Waterloo, Waterloo, Ontario, Canada, N2L 3G1}
\affiliation{$^3$Perimeter Institute for Theoretical Physics, 31 Caroline St. N, Waterloo, Ontario, Canada, N2L 2Y5}
\affiliation{$^4$Max Planck Institute for Intelligent Systems, Spemannstra{\ss}e 38, 72076 T{\"u}bingen, Germany}
\affiliation{$^\ast$These authors contributed equally to this work.}
\date{\today}

\begin{abstract}
The problem of using observed correlations to infer causal relations is relevant to a wide variety of scientific disciplines.  Yet given correlations between just two classical variables, it is impossible to determine whether they arose from a causal influence of one on the other or a common cause influencing both, unless one can implement a randomized intervention. 
We here consider the problem of causal inference for quantum variables.  We introduce \emph{causal tomography}, which unifies and generalizes conventional quantum tomography schemes to provide a complete solution to the causal inference problem using a quantum analogue of a randomized trial. We furthermore show that, in contrast to the classical case, observed quantum correlations alone can sometimes provide a solution. We implement a quantum-optical experiment that allows us to control the causal relation between two optical modes, and two measurement schemes---one with and one without randomization---that extract this relation from the observed correlations. Our results show that entanglement and coherence, known to be central to quantum information processing, also provide a quantum advantage for causal inference. 
\end{abstract}


\maketitle


``Correlation does not imply causation.'' This slogan is meant to capture the following fact: 
it is possible to explain any joint probability distribution over two variables not only by a direct causal influence of one variable on the other, but also by a common cause acting on both.  	
We here address the question of whether a similar ambiguity holds for systems that exhibit quantum effects.  
We find that, surprisingly, it does not.  

Finding causal explanations of observed correlations is a fundamental problem in science, with applications ranging from medicine and genetics to economics~\cite{Pearl_book,Spirtes_book}. As a practical illustration, consider a drug trial. Na\"{i}vely, a correlation between the variables \emph{treatment} and \emph{recovery} may suggest a direct causal influence of the former on the latter. But suppose men are more likely than women to seek treatment, and also more likely to recover spontaneously, regardless of treatment.  In this case, gender is a common cause, inducing correlations between treatment and recovery even if there is no direct causal influence.

In order to distinguish between the two possibilities, one must replace \emph{passive observation} of the early variable with an \emph{intervention} upon it.
For instance, pharmaceutical companies  do not leave the choice of treatment to the subjects of their trials, but carefully randomize the assignment of drug or placebo. 
This ensures that the treatment variable is statistically independent of any potential common causes with recovery.  Consequently, any correlations with recovery that persist must be due to a direct causal influence. 
The question of whether there \emph{were} in fact potential common causes can also be answered if one also records whether or not there is a correlation between recovery and the subject's preferred choice of treatment.   
Thus, the ability to intervene allows for a complete solution of the causal inference problem: it reveals both which variables are causes of which others and, via the strength of the correlations, the precise mathematical form of the causal dependencies.

In this article, we consider the quantum version of this causal inference problem. The challenge is to infer, based on probing the correlations between two temporally ordered quantum systems, whether these correlations are due to a direct causal influence of one system on the other, a common cause acting on both, or a combination of the two possibilities.
An additional complication relative to the classical version of the problem is that quantum theory places restrictions on gathering information about systems; for instance, not all observables that can be defined on a system can be measured precisely at the same time.
Nonetheless, we show that the ability to intervene on the early quantum system allows for a complete solution.
This constitutes a new type of tomography, which subsumes tomography of bipartite states and tomography of processes, and promises applications for determining whether the state evolution implemented by a given device is Markovian. We implement this new type of tomography experimentally and obtain a complete description of the causal structure.

The real surprise, however, is that even if one only has the ability to \emph{passively observe} the early system, the quantum correlations hold signatures of the causal structure---in other words, certain types of correlation \emph{do} imply causation.
In a recent paper, Fitzsimons, Jones and Vedral \cite{Fitzsimons2013} defined a function of the observed correlations which acts as a witness of direct causal influence, by ruling out a purely common-cause explanation.  We here present the larger framework that places this result on an equal footing with an analogous result for common-cause relations.  Differences in the patterns of correlations generated by different causal structures were also pointed out recently in \cite{Johnson2014}, in the context of extension problems.  
We here exploit the distinctive properties of quantum correlations 
to devise, for a particular class of causal scenarios, a \emph{complete solution} of the causal inference problem using passive observation alone---a task that is impossible classically. 
We implement a family of such scenarios experimentally and show that passive observation is indeed sufficient for solving the causal inference problem in this case.

\begin{figure}[h!]
\begin{centering}
\includegraphics
{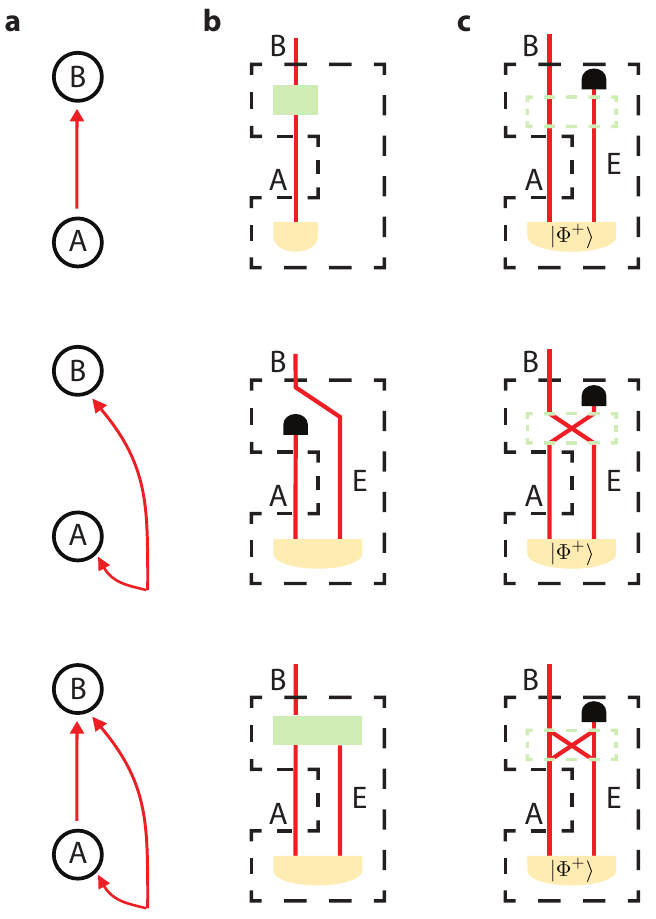}
\par\end{centering}
\caption{
{\bf The quantum causal inference problem.}
We aim to discriminate the three possible causal relations that may hold between a pair of temporally ordered quantum systems: (top to bottom) direct-cause, common-cause or a combination of both.     
(a) Directed acyclic graphs, where nodes represent quantum systems and directed edges represent causal influences, are the conventional depiction of causal structure in the causal inference literature~\cite{Pearl_book,Spirtes_book}.
(b) Quantum circuits implementing these causal structures, where wires represent quantum systems, and boxes represent operations: gates (green), state preparations (orange) and the operation of discarding the system (black). 
(c) An example of a family of quantum circuits that range over the three possible causal relations.  The gate acting on $A$ and $E$ (dashed green box) is either identity (top), swap (middle) or a probabilistic mixture of the two (bottom). }
\label{structures}
\end{figure}

{\bf The quantum causal inference problem.}
The two quantum systems whose causal structure we are probing will be denoted $A$ and $B$, with $A$ preceding $B$ in time.  The dynamics relating them may be arbitrarily complicated, involving any number of additional systems and any pattern of interactions among these.  Nonetheless, any nontrivial causal relation that is induced between $A$ and $B$ takes one of three forms:  $A$ could be a direct cause
of $B$, the two could be influenced by a common cause, or there could be a mixture of the two causal mechanisms (either a probabilistic mixture or a case where both act simultaneously).
The three possibilities are depicted in Fig.~\ref{structures}a as directed acyclic graphs and in Fig.~\ref{structures}b as quantum circuits.

A complete solution of the causal inference problem specifies not just the causal structure but also the functional relationship that holds between each system and its causal parents.  
For instance, this can be achieved by specifying the identity of the gates in the circuits depicted in Fig.~\ref{structures}b.  More generally, we aim to specify the functionality of the unknown circuit fragment that relates $A$ and $B$ (the dashed region in Fig.~\ref{structures}b).

A particular example of our causal inference problem is depicted in Fig.~\ref{structures}c.  A qubit $A$ is prepared in a maximally entangled state with an ancillary qubit $E$.  Subsequently, $A$ and $E$ are subjected to an unknown quantum operation drawn from a 1-parameter family: a probabilistic mixture of identity, with probability $1-p$, and swap, with probability $p$.  The case of pure identity corresponds to a purely direct-cause connection  between $A$ and $B$ (top), the case of pure swap corresponds to a purely common-cause connection (middle), and every other case corresponds to a hybrid of the two causal structures (bottom).  


\begin{figure}
 \centering
 \includegraphics
{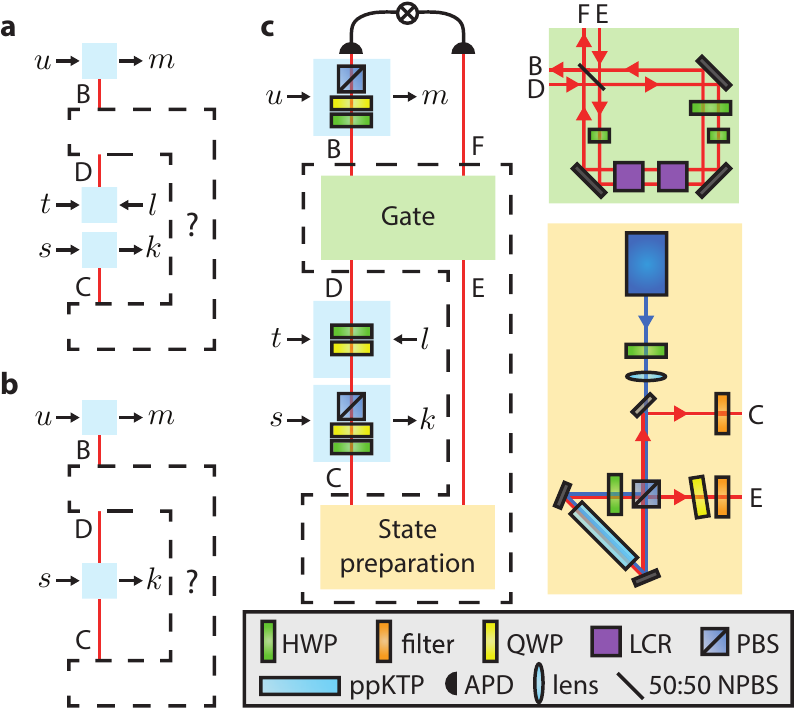}
 \caption{
 {\bf Two schemes for probing causal relations and experimental setup.}
 The unknown circuit fragment enclosed in the dashed box can be probed by two schemes.
(a) Interventionist scheme.  
The outputs $B$ and $C$ are both subjected to tomographically complete sets of measurements, while $D$ is prepared in states drawn from a tomographically complete set.  Lowercase variables denote settings and outcomes of these interventions. 
(b) Passive observation scheme.  
The outputs $B$ and $C$ are both subjected to tomographically complete sets of measurements.  The measurement on $C$ is projective, fixing the preparation on $D$.
(c) Experimental setup including polarization-entangled photon source and probabilistic swap gate.   
Notation for optical elements: half-wave plate (HWP), quarter-wave plate (QWP),  liquid crystal retarder (LCR), polarising beamsplitter (PBS), periodically-poled KTP crystal (ppKTP), avalanche photodiode (APD), and non-polarising beamsplitter (NPBS). }
 \label{setup}
\end{figure}


{\bf Intervention versus passive observation in the quantum realm.}  
The data upon which causal inference will be based is a set of correlations between the outcomes of measurements on $A$ and on $B$.  
In an interventionist scheme (depicted in Fig.~\ref{setup}a), $A$ is first measured and then reprepared in a state selected at random. In a passive observation scheme (Fig.~\ref{setup}b), system $A$ is left in the state found in the measurement, that is, it is updated according to the standard projection postulate.
It is useful to adopt a distinct notation for the versions of $A$ before and after the measurement; we denote these by $C$ and $D$ respectively. (The notational convention is natural because if $B$ is correlated with $C$, it is through a \emph{common} cause, whereas  if it is correlated with $D$, it is through \emph{direct} cause.) This trick of ``splitting" a system in order to determine its causal connection to others has also been used in the classical context \cite{Richardson2013}.

The distinction between the two quantum schemes for probing $A$ mirrors the distinction between the two classical schemes in the sense that is relevant for causal inference: while intervention provides independent information about $C$ and $D$,  passive observation provides the same information about $C$ as it does about $D$.

We will show that the quantum causal inference problem can be completely solved in the interventionist scheme, by performing informationally complete sets of measurements on $B$ and on $C$ and preparations on $D$.  
In the passive observation scheme, on the other hand, we are limited to performing an informationally complete set of measurements on $B$ and on $C$, while the preparation of $D$ is determined by the outcome of the measurement on $C$.  

We restrict ourselves to the case where $A$ and $B$ are qubits and use the Pauli observables 
and Pauli eigenstates as informationally complete sets.
Each measurement is described by two classical variables: the setting, drawn from $\{1,2,3\}$ and specifying which Pauli observable is measured, and the outcome, drawn from $\{\pm 1\}$.  We denote these by $s$ and $k$ respectively for the measurement on $C$, and by $u$ and $m$ for $B$, as depicted in Figs.~2a and 2b.  

In the interventionist scheme (Fig.~\ref{setup}a), system $D$ is prepared in the $l \in \{\pm1\}$ eigenstate of the $t\in \{1,2,3\}$ Pauli observable.  Therefore, the experimental data available for causal inference in the interventionist scheme can be represented by the conditional probability distribution $P(km|lstu)$. 

For the passive observation scheme, we measure $C$ using the standard quantum state update rule.  This can be equivalently understood as a repreparation of $D$ wherein the values of $t$ and $l$ are restricted to be equal to $s$ and $k$ respectively.
It follows that the experimental data for causal inference in this case is the conditional probability distribution $P(km|su)$.


{\bf Experiment.}
We implement the one-parameter family of circuits introduced in Fig.~\ref{structures}c, which ranges through the possible causal structures as we vary the parameter $p$, using the experimental setup shown in Fig.~\ref{setup}c.  The polarization degrees of freedom of a pair of photons constitute the pair of qubits.  We use downconversion to create entangled photon pairs in the state $\ket{\Phi^+}=\frac{1}{\sqrt{2}}(| H\rangle | H\rangle + | V\rangle | V\rangle)$, where $| H\rangle$ ($| V\rangle$) denotes horizontal (vertical) polarization. One of the photons, $C$, is subjected to a polarisation measurement, followed by a repreparation, which yields $D$. The pair of photons is then subjected to a probabilistic swap gate: with probability $p$ the modes are exchanged; otherwise they are unaffected. The first photon of the output, $B$, is subjected to a final measurement of its polarisation before both photons are detected in coincidence.   


{\bf Mathematical representation of the unknown circuit fragment.}
The dashed box in Fig.~\ref{setup}a takes one input, $D$, and produces outputs $B$ and $C$. It can therefore be represented by a
completely positive and trace-preserving (CPTP) map of the form $\mathcal{E}_{CB|D}:\mathcal{L}(\mathcal{H}_D) \to \mathcal{L}(\mathcal{H}_C\otimes \mathcal{H}_B)$, where $\mathcal{L}(\mathcal{H})$ is the space of linear operators on the Hilbert space $\mathcal{H}$.
Note, however, that the output $C$ precedes the input $D$ in time. The map $\mathcal{E}_{CB|D}$ must therefore satisfy the additional constraint that $C$ cannot depend on $D$ in any way. We term such an object a
\emph{causal map}.

Circuit fragments that do not fall into one of the standard classes (preparations, channels or measurements) have been studied in the context of alternative formulations of quantum theory by a number of authors. 
All such proposals have been motivated at least in part by the goal of describing causal structure in quantum theory, and many provide a means of describing the type of circuit fragment that we study.  
In the quantum combs framework of Chiribella, D'Ariano and Perinotti~\cite{Pavia}, $\mathcal{E}_{CB|D}$ is a particular type of 2-comb.  In the operator tensor formalism of Hardy~\cite{Hardy}, $\mathcal{E}_{CB|D}$ is an instance of an operator tensor. In the framework of Oreshkov, Costa and Brukner~\cite{Oreshkov2012}, $\mathcal{E}_{CB|D}$ is a particular type of process matrix (one that respects a global causal order). 
The ``multi-time'' formalism of Aharonov \textit{et al.}~\cite{multitime} and the general boundary formalism of Robert Oeckl~\cite{Oeckl} both have objects that would likely be able to describe the causal map $\mathcal{E}_{CB|D}$ were they suitably generalized. It is also possible to understand the causal map $\mathcal{E}_{CB|D}$ as a generalization of the notion of a quantum conditional state in the framework of Leifer and Spekkens~\cite{LeiferSpekkens}, which built on earlier work by Leifer \cite{Leifer2006}.  Indeed, it was the latter framework, with its strong connection to the field of causal inference,  that served as the primary motivation for the present work.

Note that the causal map $\mathcal{E}_{CB|D}$ incorporates as special cases both bipartite states and unipartite processes.  If the structure is purely common-cause (as in Fig.~\ref{structures}b middle), the map will have the form 
\begin{equation}
\label{eq:mapcc}
\mathcal{E}^{\rm cc}_{CB|D}= \rho_{CB} \otimes {\rm Tr}_D,
\end{equation}
describing a state on $CB$ and the trace operation on $D$.
Conversely, if the structure is purely direct-cause (as in Fig.~\ref{structures}b top), the map will have the form 
\begin{equation}
\label{eq:mapdc}
 \mathcal{E}^{\rm dc}_{CB|D}= \rho_{C} \otimes \mathcal{E}'_{B|D}, 
 \end{equation}
 describing a quantum channel from $D$ to $B$ and a normalized state on $C$.

More generally, the causal map can describe objects that are neither states nor processes (as in Fig.~\ref{structures}b bottom).  For example, if the structure is a probabilistic mixture of common-cause and direct-cause, we can write
\begin{equation}
\label{eq:map}
\mathcal{E}_{CB|D} = p \mathcal{E}^{\rm cc}_{CB|D} +(1-p) \mathcal{E}^{\rm dc}_{CB|D},
\end{equation}
where the mixing parameter $p \in [0,1]$ interpolates between the extreme cases given in Eqs.~\eqref{eq:mapcc} and \eqref{eq:mapdc}.  
An even more general form arises if direct-cause and common-cause contributions act at the same time, for instance, if the example of Fig.~\ref{structures}c is modified to allow a family of unitaries that \emph{coherently} interpolate between identity and swap.


\begin{figure*} 
  \centering
  \includegraphics
{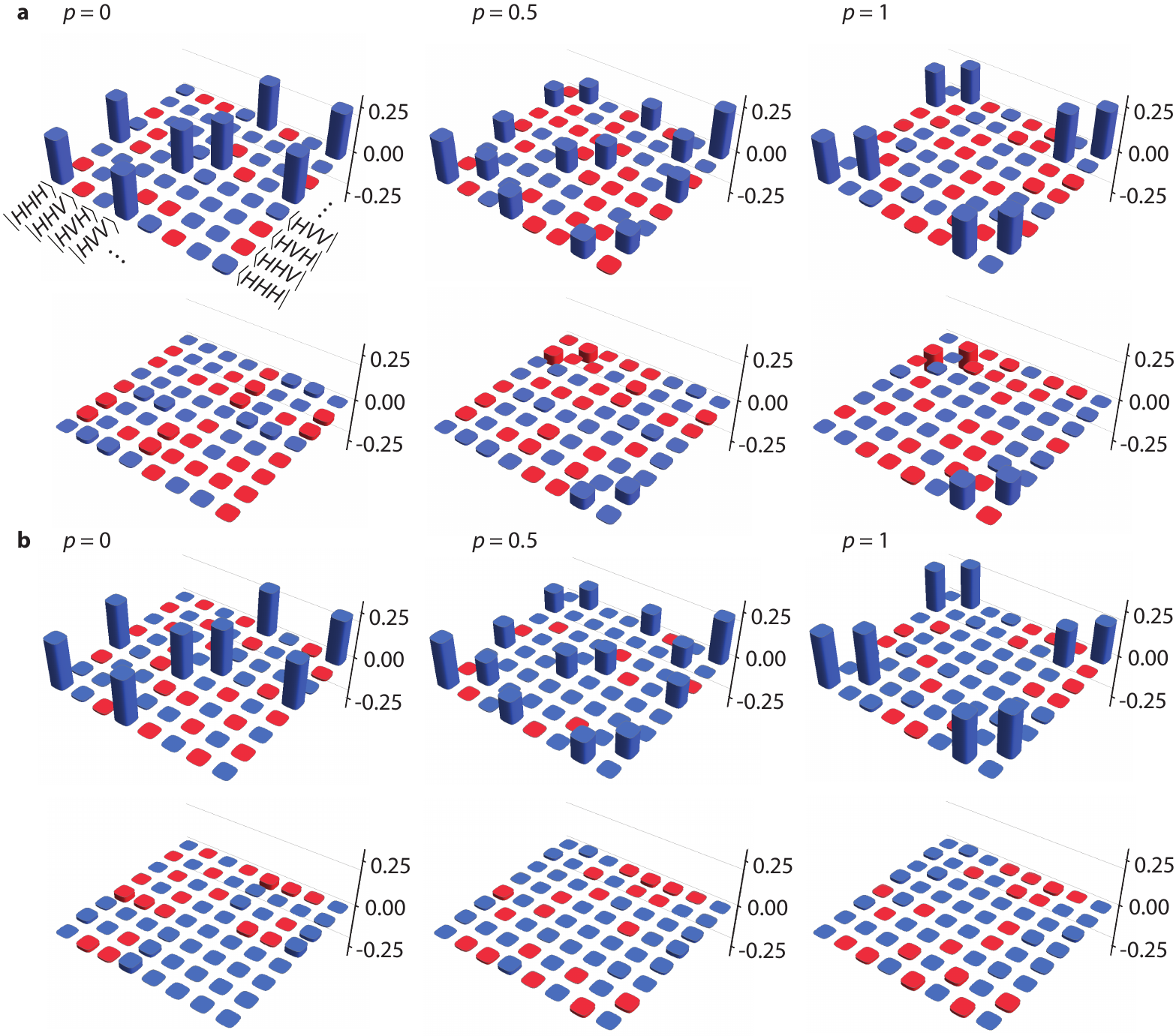} 
  \caption{\footnotesize{ 
{\bf Reconstruction of the causal map.}
Reconstructions based on (a) the interventionist scheme  and (b) the passive observation scheme, for three different causal structures (shown in Fig.~\ref{structures}c), with probability of common cause $p=0$ (left), $p=1/2$ (middle) and $p=1$ (right).
We show the Choi representation of the maps~\cite{Choi1975, DAriano2001, Altepeter2003}, which is defined as the tripartite state
$\tau_{CBD} \equiv  \left( \mathcal{E}_{CB|D^{\prime}} \otimes \mathbb{1}_{D} \right) \left( |\Phi^+ \rangle_{D'D} \langle \Phi^+|  \right)$. 
The arrays represent the real (top) and imaginary (bottom) components of the density matrices, with blue representing positive values and red negative ones. 
The basis is ordered as $CBD$, i.e. $|HHH\rangle = |H\rangle_C | H \rangle_B | H\rangle_D$.    
In the case $p=0$, we expect a Choi state $\tau_{CBD}=\frac{1}{2}\mathbb{1}_C \otimes |\Phi^{+}\rangle_{BD}\langle\Phi^{+}|$, corresponding to the map $\mathcal{E}_{CB|D}= \frac{1}{2}\mathbb{1}_{C} \otimes \mathcal{I}_{B|D}$ where $\mathcal{I}_{B|D}$ is the identity map.
The tomographically reconstructed states, denoted $\tau_{\rm fit}$, match this expectation with a fidelity 
$F \equiv {\rm Tr} \sqrt{\tau^{1/2} \tau_{\rm fit} \tau^{1/2}}$ 
(see \cite{Jozsa1994}) of $98.3\%$ using the interventionist scheme, while passive observation achieves $98.4\%$.
In the case $p=1$, we expect  $\mathcal{E}_{CB|D}= |\Phi^+\rangle_{CB}\langle \Phi^+ | {\rm Tr}_D$, which has Choi state $\tau_{CBD}=|\Phi^{+}\rangle_{CB}\langle\Phi^{+}| \otimes \frac{1}{2}{\mathbb 1}_D$.  The tomographically reconstructed states have fidelity (a) $89.9\%$ and (b) $98.4\%$ with this expected state.  We note that the data from the interventionist scheme is best fit by a state on $CB$ that deviates slightly from $ |\Phi^+\rangle$, suggesting that the state prepared experimentally did not quite match what we sought to prepare, a fact that is revealed by our causal tomography scheme.
Finally, in the case $p=1/2$, we expect the equal mixture of the two previous cases, which we find with fidelity (a) $95.2\%$ and (b) $96.0\%$.
} }
 \label{choistates}
\end{figure*}


{\bf Data analysis in the interventionist scheme.}
The conditional probability distribution $P(km|lstu)$ obtained in the interventionist scheme is sufficient to tomographically reconstruct the map $\mathcal{E}_{CB|D}$. This is proven in section~\ref{causaltomo} in the appendix. The key is that the set of preparations on $D$ span $\mathcal{L}(\mathcal{H}_D)$, the real vector space of linear operators on the Hilbert space of $D$, and the sets of measurements on $B$ and $C$ span $\mathcal{L}(\mathcal{H}_B)$ and $\mathcal{L}(\mathcal{H}_C)$ respectively, so that together they completely characterize the input-output functionality of the map, in the same way that informationally complete sets of preparations and measurements allow conventional tomography of states and processes. 
We term this scheme \emph{causal tomography} since it achieves a complete solution of the causal inference problem. 
Considering that the map $\mathcal{E}_{CB|D}$ subsumes bipartite states and processes as special cases, but also describes more general possibilities, causal tomography constitutes a novel, more general scheme that includes conventional tomography as limiting cases.

We apply our scheme to tomographically reconstruct $\mathcal{E}_{CB|D}$ from data obtained in the experiment that implements the interventionist scheme. The resulting maps are presented in Fig.~\ref{choistates}a and are found to achieve an average fidelity of $94.5\%$ with the maps that we sought to implement. 

\begin{figure}
  \centering
  \includegraphics
{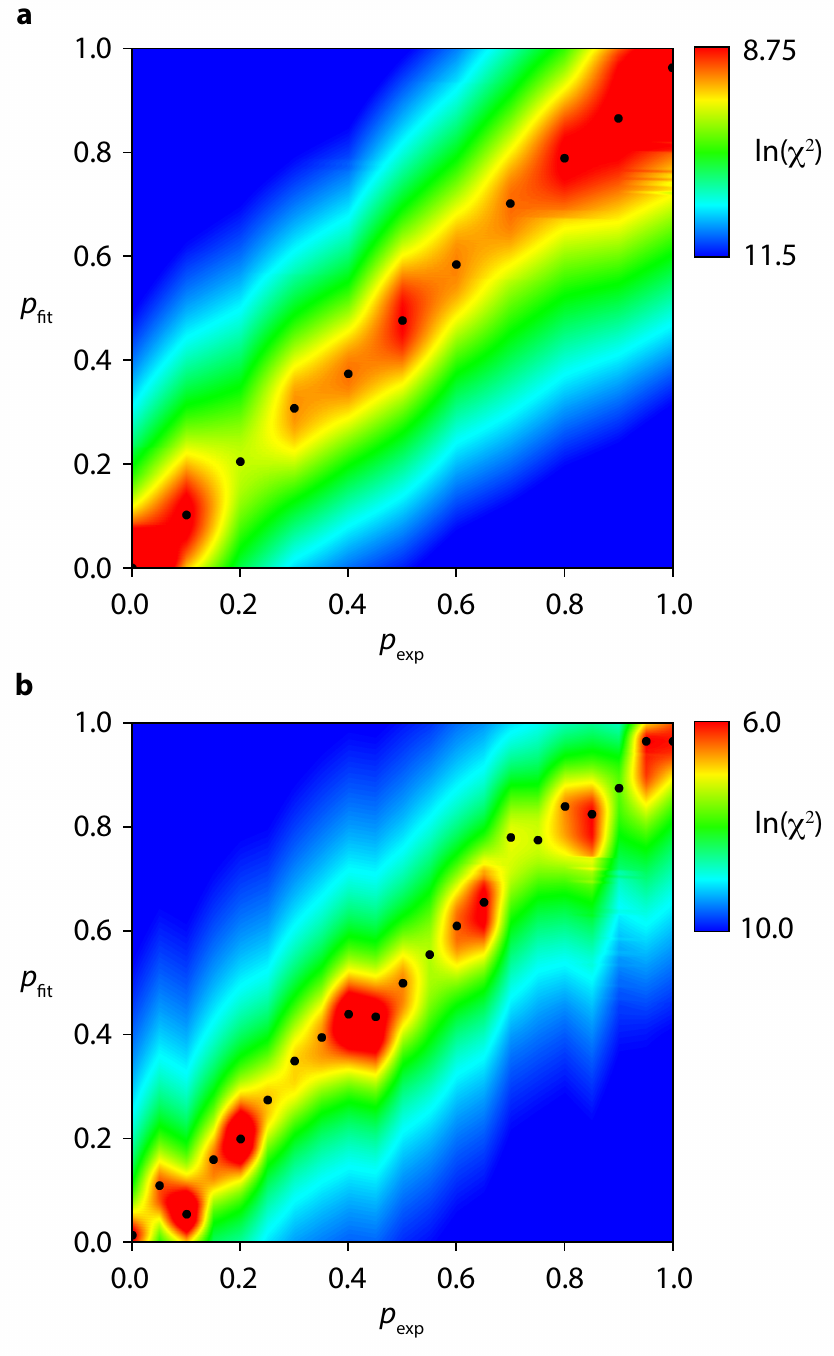} 

  \caption{\footnotesize{ {\bf Indicators of causal structure determined by interventions and passive observation.} 
We probe a probabilistic mixture of common-cause, with probability $p_{\rm{exp}}$, and direct-cause, using (a) the interventionist scheme or (b) passive observation, and fit to a mixed causal structure with probability of common-cause $p_{\rm{fit}}$. 
Colour encodes the quality of the fit, as measured by the logarithm of the least-squares residue $\chi^{2}$. 
The narrow valley of good fit around $p_{\rm{fit}}=p_{\rm{exp}}$ shows that our analysis recovers the correct value for the probability, thereby identifying the causal structure, with a root mean square (rms) deviation of $0.024$ in the interventionist scheme and $0.032$ using passive observation.}}
  \label{pfitvspexp}
\end{figure}

Although the reconstructed map constitutes a complete description of the causal mechanism, one is sometimes interested in more coarse-grained information.
For instance, for the case of a probabilistic mixture of direct-cause and common-cause mechanisms, learning the value of the mixing parameter $p$ is sufficient to determine the causal structure, if not the exact functional relationships.
To estimate $p$ in the interventionist scheme, we fit the experimental data to a map of the form of Eq.~\eqref{eq:map}.
Fig.~\ref{pfitvspexp}a shows our best estimate of $p$ as a function of the value that the experiment sought to implement. Our scheme is shown to extract $p$ with high accuracy, with an rms deviation  from the implemented value of only $0.024$.


{\bf Data analysis in the passive observation scheme.} 
Unlike the interventionist scheme, the passive observation scheme does not allow a tomographic reconstruction of the map $\mathcal{E}_{CB|D}$ in an arbitrary causal scenario.
This is because, without the randomizing repreparation, the state prepared on $D$ is the same one found in the measurement on $C$. Therefore, although the measurements on $C$ span the operator space $\mathcal{L}(\mathcal{H}_C)$ and the repreparations of $D$ span $\mathcal{L}(\mathcal{H}_D)$,  they do not together span the operator space $\mathcal{L}(\mathcal{H}_C\otimes \mathcal{H}_D)$.

Nonetheless, the correlations that are observed between $A$ and $B$ in the passive observation scheme may still contain a signature of the causal structure, as demonstrated by the simple examples in Table~1.
Indeed, it turns out that one can perfectly distinguish any unitary process from any pure maximally entangled bipartite state. If the causal scenario is a probabilistic mixture of these two possibilities, then passive observation allows one to infer both the probability and the exact nature of the process and the bipartite state, thereby affording a complete solution of the causal inference problem (up to one choice of sign). An explicit proof of the possibility of such an inference is provided in section~\ref{promise} of the appendix. 

Our experiment implements such a mixture, with the process and the state chosen such as to remove the aforementioned ambiguity, and indeed we obtain a complete solution of the causal inference problem.
Fig.~\ref{choistates}b displays reconstructions of causal maps from data obtained in the passive observation scheme, which achieve an average fidelity of $97.6\%$, on par with the results from the interventionist scheme. 
In Fig.~\ref{pfitvspexp}b, we plot our best fit for the mixing parameter $p$ based on the data from passive observation, finding an rms deviation of $0.032$ from the implemented values, again comparable to what was obtained using the interventionist scheme.
Considering that in a classical context, causal inference about a pair of variables is not possible based on passive observations alone, our results demonstrate a quantum advantage for this problem. 


\begin{table}[h]
\begin{tabular}{|c|c|c||c|| c|c|}
\hline
\multicolumn{4}{|c||} {\parbox[c][0.9cm]{3.0cm}{Pattern of correlations} }& \parbox[c]{2.2cm}{Direct-cause explanation?} & \parbox[c]{2.2cm}{Common-cause explanation?} \\
\hline
$C_{11}$ & $C_{22}$ & $C_{33}$ & $\Pi_{s}\;C_{ss}$&&\\
\hline
$+1$ & $+1$ & $+1$ & $+1$& $\mathcal{E}(\cdot) = \mathbb{1}(\cdot)\mathbb{1} $ & No \\
$+1$ & $-1$ & $-1$ & $+1$& $\mathcal{E}(\cdot) = \sigma_1(\cdot)\sigma_1 $ & No \\
$-1$ & $+1$ & $-1$ & $+1$& $\mathcal{E}(\cdot) = \sigma_2(\cdot)\sigma_2 $  & No \\
$-1$ & $-1$ & $+1$ & $+1$& $\mathcal{E}(\cdot) =  \sigma_3(\cdot)\sigma_3 $ & No \\
\hline
$-1$ & $-1$ & $-1$ & $-1$& No &$\rho = |\Psi^- \rangle \langle \Psi^-|$ \\
$-1$ & $+1$ & $+1$ & $-1$& No & $\rho = |\Phi^- \rangle \langle \Phi^-|$ \\
$+1$ & $-1$ & $+1$ & $-1$& No & $\rho = |\Phi^+ \rangle \langle \Phi^+|$ \\
$+1$ & $+1$ & $-1$ & $-1$& No & $\rho = |\Psi^+ \rangle \langle \Psi^+|$  \\
\hline 
\end{tabular}
\caption{ {\bf Signatures of causal structure accessible by passive observation.}  
Suppose the same Pauli observable is measured on both $A$ and $B$, i.e., 
$(s,u) \in \{(1,1),(2,2),(3,3)\}$, and suppose the outcomes $k$ and $m$ are found to be perfectly correlated, either positively or negatively, with correlation coefficients $C_{su}\equiv p(k=m|su)- p(k\ne m|su)\in \{ \pm1\}$.
For simplicity, we also assume that the marginal distribution over $k$ (respectively $m$) is uniform for all values $s$ (respectively $u$).
In this case, perfect negative correlation for all three observables can only be explained by a common cause, namely, when the measurements are implemented on two qubits prepared in the singlet state $\rho = |\Psi^- \rangle \langle \Psi^-|$.
There is no channel that produces this pattern: it would constitute a universal NOT gate, which is not a completely positive map.  
Similarly, positive correlation for all three observables admits of a direct-cause explanation,
namely, when the measurements are implemented on the input and output of the identity channel $\mathcal{E}(\cdot) = \mathbb{1}(\cdot)\mathbb{1}$. No bipartite state has this pattern of correlations, a fact sometimes described as the nonexistence of an `antisinglet' state. Every row of the table can be explained in this fashion. 
It emerges that it is the product of the correlation coefficients, $C_{11}\cdot C_{22} \cdot C_{33}$, which contains the signature of the causal structure.
Notation: $\sigma_1$, $\sigma_2$ and $\sigma_3$ are the Pauli matrices and $|\Psi^{\pm}\rangle \equiv \frac{1}{\sqrt{2}} (|0 \rangle |1\rangle \pm |1 \rangle |0\rangle)$, $|\Phi^{\pm}\rangle \equiv \frac{1}{\sqrt{2}} (|0 \rangle |0\rangle \pm |1 \rangle |1\rangle)$ are the Bell states, with $\{ |0\rangle, |1\rangle \}$ the eigenstates of $\sigma_3$.
 }
\end{table}


\vspace{0.5cm}
\section*{Discussion}

The example from Table I makes use of entangled states and coherent channels. This is not an accident.  
Common-cause mechanisms that prepare separable states and direct-cause mechanisms that implement entanglement-breaking channels (`measure and reprepare') produce the same patterns of correlations under passive observation
and therefore it is not possible to determine the causal structure in these cases.
We conclude that entanglement and coherence are necessary for achieving the quantum advantage in causal inference. In appendix~\ref{coherence}, we show that these conditions are also sufficient if we are promised either a purely common-cause or a purely direct-cause relation between two qubits.

The causal inference schemes described here promise extensive applications in experiments exhibiting quantum effects.  For instance, they can provide a test of whether the dynamics of a given open quantum system is Markovian or not \cite{Wolf2008, Laine2010,  Rivas2014, Rivas2010, Lu2010,  Liu2011, Tang2012, Wallman2014}.  This is because in a non-Markovian evolution, the environment acts as a common-cause between the dynamical system at one time and the same system at a later time.  Our inference schemes may also help to detect initial correlations between system and environment, which, if unaccounted for, can lead to errors in the characterization of processes \cite{Pechukas1994, Altepeter2003, Boulant2004, Weinstein2004, Howard2006, Carteret2008}.

Our results suggest several interesting avenues for future research. What can be inferred about the map $\mathcal{E}_{CB|D}$  from passive observations alone in the case of general causal mechanisms?
What can be inferred from measurements that interpolate between passive observation and intervention?
How does one generalize such schemes from pairs of systems to arbitrary numbers of systems?

\section*{Methods}

We produce polarisation-entangled photons using parametric downconversion in a nonlinear crystal embedded in a Sagnac interferometer \cite{Kim2006,Fedrizzi2007,Biggerstaff2009}. A 10-mW laser with centre wavelength 405 nm propagates through a polarising beamsplitter (PBS), splitting into two components that travel in opposite directions in the Sagnac interferometer. Each component produces degenerate type-II phase-matched downconverted pairs at 809.5 nm in a 10-mm periodically-poled KTP (ppKTP) crystal. There is a half-wave plate (HWP) placed on one side of the crystal so that each component enters the crystal with horizontal polarisation. Upon exiting the interferometer through the PBS, the photon pair is entangled in polarisation. The exact entangled state can be set using a HWP in the pump beam and a quarter-wave plate (QWP) in one of the exiting photon paths; we prepare the maximally entangled state $\ket{\Phi^+}$. In one arm, we separate the pump from the downconverted light using a dichroic mirror. The photons are coupled into single-mode fibres after passing through a bandpass filter to reduce background.  We use quantum state tomography to characterise the source and  find an average fidelity of  98.5\% with $\ket{\Phi^+}$.

Our measurement set consists of horizontal $\ket{H}$ and vertical $\ket{V}$ polarisation, $\ket{D}=\frac{1}{\sqrt{2}}(\ket{H}+\ket{V})$, $\ket{A}=\frac{1}{\sqrt{2}}(\ket{H}-\ket{V})$, $\ket{R}=\frac{1}{\sqrt{2}}(\ket{H}+i\ket{V})$, 
and $\ket{L}=\frac{1}{\sqrt{2}}(\ket{H}- i\ket{V})$. We measure polarisation using a polarising beamsplitter (PBS) preceded by a half-wave plate (HWP) and a quarter-wave plate (QWP), which are adjusted so that only one particular eigenstate can pass. By alternating the settings of the wave plates to transmit either one or the other eigenstate of a given Pauli observable in different runs of the experiment, we obtain the same statistics as if we measured in a completely non-destructive manner, which would extract the eigenvalue of the desired Pauli observable while leaving the photon intact.


The experiment requires a gate $\mathcal{G}$ that can faithfully transmit the photon polarisations directly from $D \rightarrow B$ and $E \rightarrow F$ with probability $1-p$ and swap the photon polarisations from $D \rightarrow F$ and $E \rightarrow B$ with probability $p$. We implement this with the displaced Sagnac interferometer shown in Fig.~\ref{setup}c \cite{Kwiat2000, Nagata2007}. There are two distinct paths in the interferometer: one travelling clockwise, the other travelling anticlockwise. If there is no phase difference between the two paths, the light exits the interferometer at the same side of the beamsplitter at which it entered, with a transverse displacement; if there is a $\pi$ phase difference, it exits at the opposite side. If light is incident on both input ports of the interferometer, the zero phase shift implements the identity, whilst the $\pi$ phase shift implements the swap.

This probabilistic switching is implemented using a variable liquid crystal retarder (LCR), whose birefringence can be controlled by an external voltage. A second LCR is included and set to perform the identity operation for compensation. Three half-wave plates at $45^\circ$ are inserted in the interferometer. The clockwise path encounters the LCR after passing through both waveplates, while the anticlockwise path encounters the LCR \textit{between} the two waveplates. This asymmetry results in the birefringence affecting the two paths differently. When the LCR implements the identity, $\mathbb{1}$, both paths pick up the same phase shift, so the net effect of the gate is the identity. When it implements the phase gate, $Z$, one path picks up a $\pi$ phase shift with respect to the other, so the net effect of the gate is the swap. We switch between these two levels of birefringence probabilistically using the random number generator in LabView at a rate of 5 Hz, effectively changing between the identity and swap operations with a chosen probability $p$.

The experiments proceed as follows. After preparing the entangled state on modes $C$ and $E$, we measure the polarisation of $C$.
Assuming that the photon passes through the PBS, we can then reprepare it with another QWP and HWP in the desired state for mode $D$. Light in modes $D$ and $E$ is then sent into the gate $\mathcal{G}$. The output of the gate in mode $B$ is detected using a HWP, QWP, and PBS, and $F$ is directly detected. We detect in coincidence to ensure that the source produced the requisite state. The coincidence detection is performed using single-photon detectors and coincidence logic with a window of 3 ns. Our coincidence count rate at $B$ and $F$ is approximately 2000 Hz. 

\vspace{0.5cm}
{\bf Acknowledgments} We thank J.~M.~Donohue and J.~Lavoie for valuable discussions, and M. Mazurek for his assistance in preparing the figures.  This research was supported in part by the Natural Sciences and Engineering Research Council of Canada (NSERC), Canada Research Chairs, Industry Canada and the Canada Foundation for Innovation (CFI). Research at Perimeter Institute is supported by the Government of Canada through Industry Canada and by the Province of Ontario through the Ministry of Research and Innovation.

{\bf Author contributions} 
DJ and RWS conceived the original idea for the project. KR and RWS developed the project and the theory.
MA and KJR designed the experiment. 
MA and LV performed the experiment. 
MA, KR and KJR performed the numerical calculations.  
MA, KR, KJR and RWS analyzed the results.  
KR, MA and RWS wrote the first draft of the paper and all authors contributed to the final version.

\bibliography{Causality}

\begin{thebibliography}{39}
\expandafter\ifx\csname natexlab\endcsname\relax\def\natexlab#1{#1}\fi
\expandafter\ifx\csname url\endcsname\relax
  \def\url#1{\texttt{#1}}\fi
\expandafter\ifx\csname urlprefix\endcsname\relax\def\urlprefix{URL }\fi

\bibitem[{Pearl(2000)}]{Pearl_book}
Pearl, J.
\newblock \emph{Causality: models, reasoning and inference} (Cambridge Univ.
  Press, New York, 2000).

\bibitem[{Spirtes \emph{et~al.}(2000)Spirtes, Glymour \&
  Scheines}]{Spirtes_book}
Spirtes, P., Glymour, C. \& Scheines, R.
\newblock \emph{Causation, prediction, and search} (MIT Press, Cambridge,
  2000).

\bibitem[{Fitzsimons \emph{et~al.}(2013)Fitzsimons, Jones \&
  Vedral}]{Fitzsimons2013}
Fitzsimons, J., Jones, J. \& Vedral, V.
\newblock Quantum correlations which imply causation.
\newblock \emph{arXiv:1302.2731}  (2013).

\bibitem[{Johnson \& Viola(2014)}]{Johnson2014}
Johnson, P. \& Viola, L.
\newblock On state vs. channel quantum extension problems: exact results for
  {U}x{U}x{U} symmetry.
\newblock \emph{arXiv:1405.1062}  (2014).

\bibitem[{Richardson \& Robins(2013)}]{Richardson2013}
Richardson, T.~S. \& Robins, J.~M.
\newblock Single world internvention graphs {(SWIGs)}.
\newblock Tech. Rep. 128, CSSS, University of Washington (2013).

\bibitem[{Chiribella \emph{et~al.}(2009)Chiribella, D'Ariano \&
  Perinotti}]{Pavia}
Chiribella, G., D'Ariano, G.~M. \& Perinotti, P.
\newblock Theoretical framework for quantum networks.
\newblock \emph{Phys. Rev. A} \textbf{80}, 022339 (2009).

\bibitem[{Hardy(2012)}]{Hardy}
Hardy, L.
\newblock The operator tensor formulation of quantum theory.
\newblock \emph{Philos. T. Roy. Soc. A} \textbf{370}, 3385--3417 (2012).

\bibitem[{Oreshkov \emph{et~al.}(2012)Oreshkov, Costa \&
  Brukner}]{Oreshkov2012}
Oreshkov, O., Costa, F. \& Brukner, C.
\newblock Quantum correlations with no causal order.
\newblock \emph{Nat. Commun.} \textbf{3}, 1092 (2012).

\bibitem[{Aharonov \emph{et~al.}(2009)Aharonov, Popescu, Tollaksen \&
  Vaidman}]{multitime}
Aharonov, Y., Popescu, S., Tollaksen, J. \& Vaidman, L.
\newblock Multiple-time states and multiple-time measurements in quantum
  mechanics.
\newblock \emph{Phys. Rev. A} \textbf{79}, 052110 (2009).

\bibitem[{Oeckl(2003)}]{Oeckl}
Oeckl, R.
\newblock A ``general boundary'' formulation for quantum mechanics and quantum
  gravity.
\newblock \emph{Phys. Lett. B} \textbf{575}, 318--324 (2003).

\bibitem[{Leifer \& Spekkens(2013)}]{LeiferSpekkens}
Leifer, M. \& Spekkens, R.~W.
\newblock Towards a formulation of quantum theory as a causally neutral theory
  of Bayesian inference.
\newblock \emph{Phys. Rev. A} \textbf{88}, 052130 (2013).

\bibitem[{Leifer(2006)}]{Leifer2006}
Leifer, M.~S.
\newblock Quantum dynamics as an analog of conditional probability.
\newblock \emph{Phys. Rev. A} \textbf{74}, 042310 (2006).

\bibitem[{Choi(1975)}]{Choi1975}
Choi, M.~D.
\newblock Completely positive linear maps on complex matrices.
\newblock \emph{Linear Algebra Appl.} \textbf{10}, 285--290 (1975).

\bibitem[{D'Ariano \& Lo~Presti(2001)}]{DAriano2001}
D'Ariano, G.~M. \& Lo~Presti, P.
\newblock Quantum Tomography for Measuring Experimentally the Matrix Elements
  of an Arbitrary Quantum Operation.
\newblock \emph{Phys. Rev. Lett.} \textbf{86}, 4195--4198 (2001).

\bibitem[{Altepeter \emph{et~al.}(2003)}]{Altepeter2003}
Altepeter, J.~B. \emph{et~al.}
\newblock Ancilla-Assisted Quantum Process Tomography.
\newblock \emph{Phys. Rev. Lett.} \textbf{90}, 193601 (2003).

\bibitem[{Jozsa(1994)}]{Jozsa1994}
Jozsa, R.
\newblock Fidelity for mixed quantum states.
\newblock \emph{J. Mod. Opt.} \textbf{41}, 2315--2323 (1994).

\bibitem[{Wolf \emph{et~al.}(2008)Wolf, Eisert, Cubitt \& Cirac}]{Wolf2008}
Wolf, M.~M., Eisert, J., Cubitt, T. \& Cirac, J.
\newblock Assessing non-{M}arkovian quantum dynamics.
\newblock \emph{Phys. Rev. Lett.} \textbf{101}, 150402 (2008).

\bibitem[{Laine \emph{et~al.}(2010)Laine, Piilo \& Breuer}]{Laine2010}
Laine, E.-M., Piilo, J. \& Breuer, H.-P.
\newblock Measure for the non-Markovianity of quantum processes.
\newblock \emph{Phys. Rev. A} \textbf{81}, 062115 (2010).

\bibitem[{Rivas \emph{et~al.}(2014)Rivas, Huelga \& Plenio}]{Rivas2014}
Rivas, {\'A}., Huelga, S.~F. \& Plenio, M.~B.
\newblock Quantum Non-{M}arkovianity: Characterization, Quantification and
  Detection.
\newblock \emph{arXiv:1405.0303}  (2014).

\bibitem[{Rivas \emph{et~al.}(2010)Rivas, Huelga \& Plenio}]{Rivas2010}
Rivas, {\'A}., Huelga, S.~F. \& Plenio, M.~B.
\newblock Entanglement and non-{M}arkovianity of quantum evolutions.
\newblock \emph{Phys. Rev. Lett.} \textbf{105}, 050403 (2010).

\bibitem[{Lu \emph{et~al.}(2010)Lu, Wang \& Sun}]{Lu2010}
Lu, X.-M., Wang, X. \& Sun, C.
\newblock Quantum {F}isher information flow and non-{M}arkovian processes of
  open systems.
\newblock \emph{Phys. Rev. A} \textbf{82}, 042103 (2010).

\bibitem[{Liu \emph{et~al.}(2011)}]{Liu2011}
Liu, B.-H. \emph{et~al.}
\newblock Experimental control of the transition from {M}arkovian to
  non-{M}arkovian dynamics of open quantum systems.
\newblock \emph{Nat. Phys.} \textbf{7}, 931--934 (2011).

\bibitem[{Tang \emph{et~al.}(2012)}]{Tang2012}
Tang, J.-S. \emph{et~al.}
\newblock Measuring non-{M}arkovianity of processes with controllable
  system-environment interaction.
\newblock \emph{Europhys. Lett.} \textbf{97}, 10002 (2012).

\bibitem[{Wallman \emph{et~al.}(2014)Wallman, Flammia, Barnhill \&
  Emerson}]{Wallman2014}
Wallman, J., Flammia, S., Barnhill, M. \& Emerson, J.
\newblock Simpler, faster, better: robust randomized benchmarking tests for
  non-unitality and non-{M}arkovianity in quantum devices.
\newblock In \emph{Bulletin of the American Physical Society} (2014).

\bibitem[{Pechukas(1994)}]{Pechukas1994}
Pechukas, P.
\newblock Reduced Dynamics Need Not Be Completely Positive.
\newblock \emph{Phys. Rev. Lett.} \textbf{73}, 1060--1062 (1994).

\bibitem[{Boulant \emph{et~al.}(2004)Boulant, Emerson, Havel, Cory \&
  Furuta}]{Boulant2004}
Boulant, N., Emerson, J., Havel, T.~F., Cory, D.~G. \& Furuta, S.
\newblock Incoherent noise and quantum information processing.
\newblock \emph{J. Chem. Phys.} \textbf{121}, 2955--2961 (2004).

\bibitem[{Weinstein \emph{et~al.}(2004)}]{Weinstein2004}
Weinstein, Y.~S. \emph{et~al.}
\newblock Quantum process tomography of the quantum {F}ourier transform.
\newblock \emph{J. Chem. Phys.} \textbf{121}, 6117--6133 (2004).

\bibitem[{Howard \emph{et~al.}(2006)}]{Howard2006}
Howard, M. \emph{et~al.}
\newblock Quantum process tomography and {L}indblad estimation of a solid-state
  qubit.
\newblock \emph{New J. Phys.} \textbf{8}, 33 (2006).

\bibitem[{Carteret \emph{et~al.}(2008)Carteret, Terno \&
  Zyczkowski}]{Carteret2008}
Carteret, H., Terno, D.~R. \& Zyczkowski, K.
\newblock Physical accessibility of non-completely positive maps.
\newblock \emph{Phys. Rev. A} \textbf{77}, 042113 (2008).

\bibitem[{Kim \emph{et~al.}(2006)Kim, Fiorentino \& Wong}]{Kim2006}
Kim, T., Fiorentino, M. \& Wong, F. N.~C.
\newblock Phase-stable source of polarization-entangled photons using a
  polarization Sagnac interferometer.
\newblock \emph{Phys. Rev. A} \textbf{73}, 012316 (2006).

\bibitem[{Fedrizzi \emph{et~al.}(2007)Fedrizzi, Herbst, Poppe, Jennewein \&
  Zeilinger}]{Fedrizzi2007}
Fedrizzi, A., Herbst, T., Poppe, A., Jennewein, T. \& Zeilinger, A.
\newblock A wavelength-tunable fiber-coupled source of narrowband entangled
  photons.
\newblock \emph{Opt. Express} \textbf{15}, 15377--15386 (2007).

\bibitem[{Biggerstaff \emph{et~al.}(2009)}]{Biggerstaff2009}
Biggerstaff, D.~N. \emph{et~al.}
\newblock Cluster-State Quantum Computing Enhanced by High-Fidelity Generalized
  Measurements.
\newblock \emph{Phys. Rev. Lett.} \textbf{103}, 240504 (2009).

\bibitem[{Kwiat \emph{et~al.}(2000)Kwiat, Mitchell, Schwindt \&
  White}]{Kwiat2000}
Kwiat, P.~G., Mitchell, J.~R., Schwindt, P. D.~D. \& White, A.~G.
\newblock {G}rover's search algorithm: an optical approach.
\newblock \emph{J. Mod. Opt.} \textbf{47}, 257--266 (2000).

\bibitem[{Nagata \emph{et~al.}(2007)Nagata, Okamoto, O'Brien, Sasaki \&
  Takeuchi}]{Nagata2007}
Nagata, T., Okamoto, R., O'Brien, J.~L., Sasaki, K. \& Takeuchi, S.
\newblock Beating the Standard Quantum Limit with Four-Entangled Photons.
\newblock \emph{Science} \textbf{316}, 726--729 (2007).

\bibitem[{Bengtsson \& Zyczkowski(2006)}]{Bengtsson2006}
Bengtsson, I. \& Zyczkowski, K.
\newblock \emph{Geometry of quantum states} (Cambridge Univ. Press, New York,
  2006).

\bibitem[{Jamio{\l}kowski(1972)}]{Jamiolkowski1972}
Jamio{\l}kowski, A.
\newblock Linear transformations which preserve trace and positive
  semidefiniteness of operators.
\newblock \emph{Rep. Math. Phys.} \textbf{3}, 275--278 (1972).

\bibitem[{Wiseman \emph{et~al.}(2007)Wiseman, Jones \& Doherty}]{Wiseman2007}
Wiseman, H.~M., Jones, S.~J. \& Doherty, A.~C.
\newblock Steering, Entanglement, Nonlocality, and the
  {E}instein-{P}odolsky-{R}osen Paradox.
\newblock \emph{Phys. Rev. Lett.} \textbf{98}, 140402 (2007).

\bibitem[{Jevtic \emph{et~al.}(2013)Jevtic, Pusey, Jennings \&
  Rudolph}]{Jevtic2013}
Jevtic, S., Pusey, M.~F., Jennings, D. \& Rudolph, T.
\newblock The quantum steering ellipsoid.
\newblock \emph{arXiv:1303.4724}  (2013).

\bibitem[{James \emph{et~al.}(2001)James, Kwiat, Munro \& White}]{James2001}
James, D.~F., Kwiat, P.~G., Munro, W.~J. \& White, A.~G.
\newblock Measurement of qubits.
\newblock \emph{Phys. Rev. A} \textbf{64}, 052312 (2001).

\end{thebibliography}
\bibliographystyle{nature2}

\newpage
\begin{center}
\textbf{\large Appendix}
\end{center}
\makeatletter

\section{Interventionist scheme}


\subsection{Causal tomography}\label{causaltomo}

In this section, we show that the probability distribution obtained in the interventionist scheme, $P(km|lstu)$, completely specifies the causal map $\mathcal{E}_{CB|D}$ describing the unknown circuit fragment (dashed box in Fig.~\ref{setup}a in the main article). The limiting cases of purely common-cause and purely direct-cause relations are discussed in more detail in the next section, showing how the generic scheme reduces to tomography of bipartite states and single-system processes respectively. 

We take $s,t,u \in \left\{1,2,3\right\}$ to index the three Pauli operators $\left\{\sigma_{1},\sigma_{2},\sigma_{3}\right\} $ and by extension their eigenbases. The values $k,l,m \in \{+1,-1\}$ specify the eigenstates, and $\Pi_{sk}$, $\Pi_{tl}$, and $\Pi_{um}$ denote the projectors onto those eigenstates.
Rather than referring directly to the map $\mathcal{E}_{CB|D}$, it is convenient to introduce the Jamio\l{}kowski representation of the map~\cite{Bengtsson2006, Jamiolkowski1972, Choi1975, LeiferSpekkens}: the operator $\rho_{CB|D} \in \mathcal{L}(\mathcal{H}_C \otimes \mathcal{H}_B \otimes \mathcal{H}_D)$ defined by
\begin{equation}
\rho_{CB|D} = \rm{Tr}_{D^{\prime}} \left[ \left( \mathcal{E}_{CB|D^{\prime}} \otimes \mathbb{1}_{D} \right) \left( |\Phi^+ \rangle_{D'D} \langle \Phi^+| ^{T_D} \right) \right],
\end{equation}
where $T_D$ denotes the partial transpose on $D$. (The Choi state, $\tau_{CB|D}$, which we introduced in the main article to represent the map $\mathcal{E}_{CB|D}$, differs from $\rho_{CB|D}$ by a partial transpose on $D$ and a normalization factor.) One can express the action of the map $\mathcal{E}_{CB|D}$ on an arbitrary state $\rho_D$ in terms of $\rho_{CB|D}$ as
\begin{equation}\label{Jamio2}
\mathcal{E}_{CB|D}(\rho_{D}) =  {\rm Tr}_{D} \left[ \rho_{CB|D}( \mathbb{1}_{CB} \otimes \rho_{D} )\right].
\end{equation}

Assuming the input system $D$ is prepared in the Pauli eigenstate $\Pi_{tl}$ and $\sigma_{s}$ is measured on $C$ while $\sigma_{u}$ is measured on $B$, the probability of obtaining outcomes $k$, $m$ can be expressed in terms of the Jamio\l{}kowski operator $\rho_{CB|D}$ as
\begin{equation}
P(km|lstu)={\rm Tr}_{CBD} \left[ \rho_{CB|D}(\Pi_{sk}^{C}\otimes \Pi_{um}^{B}\otimes \Pi_{tl}^{D}) \right].
\label{tomoprobandChoi}
\end{equation}

It is convenient to work with the distribution $P(klm|stu)$ rather than $P(km|lstu)$. Given that $l$ is chosen uniformly at random from $\{+1,-1\}$, independently of $s,t$, or $u$, the relation between the two is simply \begin{equation}
P(klm|stu)=\frac{1}{2}P(km|lstu).
\label{conversion}
\end{equation}
In essence, one can think of the preparation of $D$ as a filtering-type measurement of the Pauli observable $\sigma_t$ acting on the maximally mixed state on $D$. In this scenario, it is natural to consider the joint distribution of outcomes for three measurements, $P(klm|stu)$.

Eqs.~\eqref{tomoprobandChoi} and ~\eqref{conversion} imply that from $P(klm|stu)$ one can determine the Hilbert-Schmidt inner product of $\rho_{CB|D}$ with certain elements of the operator space $\mathcal{L}(\mathcal{H}_C \otimes \mathcal{H}_B \otimes \mathcal{H}_D)$, namely, the Pauli operators, $\sigma_s^C \otimes \sigma_u^B \otimes \sigma_t^D$, for $s,t,u\in \left\{1,2,3\right\}$. 
We refer to these components of $\rho_{CB|D}$ as correlators and denote them $C_{stu}$,
\begin{eqnarray}
C_{stu} &\equiv& {\rm Tr} \left[ \rho_{CB|D}(\sigma_{s}^{C}\otimes \sigma_{u}^{B} \otimes  \sigma_{t}^{D}) \right]\nonumber \\
&=& 2 \sum_{k,l,m=\pm1}klm\; P\left(klm|stu\right).
\end{eqnarray}
To reconstruct $\rho_{CB|D}$, however, we need to know its components relative to a \emph{complete basis} of the operator space.
Such a basis is provided by products of the set of Pauli operators only if this set includes the identity operator.  Defining $\sigma_0 = \mathbb{1}$ and introducing variables $s',t',u'$ with range $\{0,1,2,3\}$,
the set of product operators $\sigma_{s'}^C \otimes \sigma_{t'}^D \otimes \sigma_{u'}^B$ provides a complete basis.  We again refer to the components in this basis as correlators, and denote them $C_{s't'u'}$,
\begin{equation}\label{correlatorsprime}
C_{s't'u'}\equiv {\rm Tr} \left[ \rho_{CB|D}(\sigma_{s'}^{C}\otimes \sigma_{u'}^{B}\otimes \sigma_{t'}^{D}) \right],
\end{equation}
where $s',t',u'\in\{0,1,2,3\}$.

The cases wherein one of $s',t',u'$ is zero while the other two are in $\{1,2,3\}$ describe correlations between Pauli observables for a pair of systems, for instance, $C_{s0u} \equiv {\rm Tr}_{CBD} \left[ \rho_{CB|D}(\sigma_{s}^{C} \otimes \sigma_{u}^{B}\otimes \mathbb{1}_D) \right]$ for $s,u \in \{1,2,3\}$.
Note that the marginal of $P(klm|stu)$ on any two outcome variables is independent of the value of the setting variable for the third: $P(kl|stu)=P(kl|st)$, $P(km|stu)=P(km|su)$, and $P(lm|stu)=P(lm|tu)$.  The only subtle case is the second one, which follows from the fact that $l$ is chosen uniformly at random, such that the state on $D$ when one is not conditioning on $l$ is independent of $t$, $\sum_l \frac{1}{2} \Pi_{t,l} =\frac{1}{2} \mathbb{1}$.
It follows that the correlators in question can be expressed as:
\begin{eqnarray}
C_{s0u} &=&  2 \sum_{k,m=\pm1}km\; P\left(km|su\right),\nonumber\\
C_{0tu} &=& 2 \sum_{l,m=\pm1}lm\; P\left(lm|tu\right),\nonumber\\
C_{st0} &=& 2 \sum_{k,l=\pm1}kl\; P\left(kl|st\right).
\end{eqnarray}
Furthermore, since $C$ precedes $D$ in time, it cannot depend on $D$. Hence $P(kl|st)=P(k|s)P(l|t)$, and, given that $l$ is chosen uniformly at random, we find that $C_{st0}=0$. 
 
The cases wherein two of $s',t',u'$ are zero and only one is in $\{1,2,3\}$ describe marginal expectations for Pauli observables, for instance, 
$C_{s00} \equiv {\rm Tr}_{CBD} \left[ \rho_{CB|D}(\sigma_{s}^{C}\otimes \mathbb{1}_B \otimes \mathbb{1}_D) \right]$ for $s \in \{1,2,3\}$. 
As before, the marginal on a single outcome variable is independent of the other two setting variables: $P(k|stu)=P(k|s)$, $P(l|stu)=P(l|t)$, and $P(m|stu)=P(m|u)$.  
Again, the only subtle case is the second one, which follows from the fact that $l$ is chosen uniformly at random. The correlators can therefore be expressed as:

\begin{eqnarray}
C_{s00} &=& 2 \sum_{k=\pm1}k\; P\left(k|s\right),\nonumber\\
C_{0t0} &=& 2 \sum_{l=\pm1}l\; P\left(l|t\right),\nonumber\\
C_{00u} &=& 2 \sum_{m=\pm1}m\; P\left(m|u\right).
\end{eqnarray}
The fact that $l$ is chosen uniformly at random also implies that $C_{0t0}=0$ for all $t$.

Finally, from the fact that $\mathcal{E}_{CB|D}$ is trace-preserving, it follows that ${\rm Tr}_{CB} \rho_{CB|D} = \mathbb{1}_D$, so that ${\rm Tr}_{CBD}\left[\rho_{CB|D}\right]=2$, and hence
\begin{equation}
C_{000} = 2. 
\end{equation}
\color{black}

Thus each of the correlators $C_{s't'u'}$ can be calculated from the measured statistics.  
Because the Pauli operators form an orthogonal basis of the operator space relative to the Hilbert-Schmidt inner product, we can invert Eq.~\eqref{correlatorsprime} and reconstruct $\rho_{CB|D}$ via
\begin{equation}\label{reconst}
\rho_{CB|D}=\frac{1}{8}\sum_{s',t',u'=0}^{3}C_{s't'u'}\sigma_{s'}^{C}\otimes\sigma_{u'}^{B}\otimes\sigma_{t'}^{D}.
\end{equation}
The map $\mathcal{E}_{CB|D}$ is then recovered from $\rho_{CB|D}$ using Eq.~\eqref{Jamio2}.

We refer to this scheme, by which the causal map $\mathcal{E}_{CB|D}$ is reconstructed from the measurement statistics, as \emph{causal tomography}.


\subsection{Special cases: tomography of processes and of bipartite states}

This section describes how causal tomography reduces to tomography of bipartite states and tomography of single-system processes in the cases where the causal structure is purely common-cause and purely direct-cause respectively.

In the case of a purely common-cause relation, neither $B$ nor $C$ depend on which state $\Pi_{lt}$ is prepared on $D$, so that
\begin{equation}
P\left(klm|stu\right)=P\left(km|su\right).
\end{equation}
It follows that there are no triple-wise correlations, $C_{stu}=0$. We already found that, since $D$ has no influence on $C$, $C_{st0}=0$ and $C_{0t0}=0$. 
Furthermore, the lack of a causal connection between $D$ and $B$ in the purely common-cause case implies $P(lm|stu)=\frac{1}{2}P(m|u)$, and consequently  $C_{0tu}=0$.  
Consequently, the only nonzero correlators for the purely common-cause scenario are those of the form $C_{s'0u'}$.  
Hence, the correlators can be expressed as $C_{s't'u'}=2 \delta_{t',0} C_{s'u'}$,  with
\begin{eqnarray}
C_{su} &\equiv& \sum_{k,m=\pm1}km\; P\left(km|su\right),\nonumber\\
C_{s0} &\equiv& \sum_{k=\pm1}k\; P\left(k|s\right), \nonumber \\
C_{0u} &\equiv& \sum_{m=\pm1}m\; P\left(m|u\right), \nonumber \\
C_{00} &\equiv& 1.
\end{eqnarray}
The reconstruction presented in Eq.~\eqref{reconst} yields
\begin{equation}
\rho_{CB|D}=\rho_{CB}\otimes \mathbb{1}_{D},
\end{equation}
with
\begin{equation}
\rho_{CB}=\frac{1}{4}\sum_{s'u'=0}^{3}C_{s'u'}\sigma_{s'}^{C}\otimes\sigma_{u'}^{B},
\end{equation}
which is precisely the expression for the tomographic reconstruction of a two-qubit state from the correlators between Pauli operators on the two qubits. 

In the limiting case of a purely direct-cause relation, $C$ is not correlated with $B$ or $D$: the probability distribution factorizes into
\begin{equation}
P\left(klm|stu\right)=P\left(k|s\right)P\left(lm|tu\right),
\end{equation}
so the correlators factor as $C_{s't'u'}=C_{s'} C_{t'u'}$, with
\begin{eqnarray}
C_{tu} &=& 2 \sum_{l,m=\pm1}lm\; P\left(lm|tu\right),\nonumber\\
C_{t0} &=& 2 \sum_{l=\pm1}l\; P\left(l|t\right),\nonumber\\
C_{0u} &=& 2 \sum_{m=\pm1}m\; P\left(m|u\right),\nonumber\\
C_{00} &=& 2,\nonumber\\
C_{s} &=& \sum_{k=\pm1}k\; P\left(k|s\right),\nonumber\\
C_{0} &=& 1.
\end{eqnarray}
Note that $C_{t0}=0$ for all $t$ because $l$ is chosen uniformly at random.

From these correlators, one reconstructs a Jamio\l{}kowski operator of the form 
\begin{equation}
\rho_{CB|D}=\rho_{C} \otimes \rho_{B|D}.
\end{equation}
where
\begin{equation}\label{rhoC}
\rho_{C}=\frac{1}{2}\sum_{s'=0}^{3}C_{s'}\sigma_{s'},
\end{equation}
and
\begin{equation}\label{BtoD}
\rho_{B|D}=\frac{1}{4}\sum_{t'u'=0}^{3}C_{t'u'}\sigma_{t'}^{D}\otimes\sigma_{u'}^{B}.
\end{equation}
Clearly, Eq.~\eqref{rhoC} is the standard expression for the tomographic reconstruction of a qubit state from the expectation values of the Pauli operators.  Meanwhile, Eq.~\eqref{BtoD} is the expression for the tomographic reconstruction of the Jamio\l{}kowski representation of a single-qubit process. Denoting the completely positive trace-preserving map associated with this process by $\mathcal{E}'_{B|D}$, its action can be expressed in terms of $\rho_{B|D}$ using the standard Jamio\l{}kowski isomorphism, $\mathcal{E}'_{B|D}(\rho_D) =  {\rm Tr}_{D} \left[ \rho_{B|D}( \mathbb{1}_{B} \otimes \rho_{D} )\right]$.

\section{Passive observation}


\subsection{Signatures of causal structure}\label{signatures}

This section considers two possible causal structures: either $D$ has a direct causal influence on $B$, or $C$ and $B$ are connected by a common cause. We derive properties of the measurement statistics that reflect the underlying causal structure.

In the passive observation scheme, we perform a projective measurement on $C$, with rank-one projectors $\Pi_{sk}^C$ indexed by the setting $s$ and the outcome $k$. The measurement obeys the projection postulate, so that the state on $D$ after a measurement of $s$ yielding $k$ is given by the same projector, $\Pi_{sk}^D$. The late system, $B$, is probed by a measurement with projectors $\Pi_{um}^B$, with settings $u$ and outcomes $m$.
We can therefore relate the joint probability distribution over measurement outcomes, $P(km|su)$, to the causal map $\mathcal{E}_{CB|D}$ by
\begin{equation}
P(km|su) = {\rm Tr}_{BC} \left[ \Pi_{um}^B \otimes \Pi_{sk}^C \mathcal{E}_{CB|D}(\Pi_{sk}^D) \right].
\end{equation}

In the case where the relation is purely common-cause, the causal map reduces to the form
\begin{equation}
\mathcal{E}^{\rm cc}_{CB|D}= \rho_{CB} \otimes {\rm Tr}_D,
\end{equation}
for some bipartite state $\rho_{CB}$.  It follows that in this case, 
\begin{equation}\label{ccstatistics}
P(km|su) = {\rm Tr}_{BC} \left( \Pi_{sk}^C \otimes \Pi_{um}^B \rho_{CB} \right),
\end{equation}
which is simply the standard expression for joint statistics obtained from measurements on a pair of quantum systems.   

Moreover, noting that the projectors that constitute the measurement sum to identity, $\sum_{m}\Pi_{um}=\mathbb{1}\;\forall u$, we find the marginal distribution over $k$ to be
\begin{equation} \label{margdistcc}
P(k|su)={\rm Tr}_{C} \left( \Pi_{sk}^C \rho_{C} \right)=P(k|s),
\end{equation}
which is simply the probability distribution that we would expect from the marginal state $\rho_C \equiv {\rm Tr}_B \rho_{CB}$. We will see below that the same marginal distribution is produced by a direct-cause structure, and can therefore not serve as an indicator of causal structure. 
Instead, we consider  $P(m|ksu)$, the conditional probability of finding outcome $m$ in a  measurement on $B$ with setting $u$ given that the measurement on $C$ with setting $s$ found outcome $k$.  This can be determined from $P(km|su)$ using the chain rule
\begin{equation}
P(km|su) = P(m|ksu)P(k|su).
\end{equation} 

In order to obtain a simple expression for the conditional probability distribution, we define the \emph{conditional quantum state}, as introduced in Ref.~\cite{LeiferSpekkens}: a bipartite operator, $\rho_{B|C} \in \mathcal{L}(\mathcal{H}_C \otimes \mathcal{H}_B)$, determined by the joint state $\rho_{CB}$ through the definition
\begin{equation}
\rho_{B|C} \equiv \left(\rho_{C}^{-\frac{1}{2}}\otimes\mathbb{1}_{B}\right)\rho_{CB}\left(\rho_{C}^{-\frac{1}{2}}\otimes\mathbb{1}_{B}\right).
\end{equation}
In terms of $\rho_{B|C}$, the conditional probability distribution can be expressed as
\begin{equation}\label{CondProbCC}
P(m|ksu)={\rm Tr}_{CB} \left(\Pi_{sk}^C \otimes \Pi_{um}^B \rho_{B|C} \right).
\end{equation}
The operator $\rho_{C|B}$ is called the \emph{acausal} conditional state in Ref.~\cite{LeiferSpekkens}, to emphasize the fact that it does not describe a (direct) causal influence. Nevertheless, it encodes a rule of inference, namely what one can infer about $B$ if one finds a certain state on $C$. 
This sort of state update rule has  been studied under the name of steering \cite{Wiseman2007} and the general form of the affine map $\mathcal{E}_{B|C}$ which is Jamio\l{}kowski isomorphic to $\rho_{B|C}$ is described in proposition V.1 of Ref.~\cite{LeiferSpekkens}.  

By its definition, $\rho_{B|C}$ clearly satisfies the quantum analogue of the law of total probability, ${\rm Tr}_C \rho_{B|C} = \mathbb{1}_C$, which corresponds to the fact that the steering map $\mathcal{E}_{B|C}$ is trace-preserving.  More importantly, like the bipartite state $\rho_{CB}$ from which it is derived, the conditional $\rho_{B|C}$ is positive semi-definite. By Choi's theorem \cite{Choi1975}, the steering map $\mathcal{E}_{B|C}$ is therefore not completely positive in general, however its composition with the partial transpose on $C$, $\mathcal{E}_{B|C} \circ T_C$, is a completely positive map. We conclude that if the statistics $P(m|ksu)$ \emph{cannot} be cast in the form of Eq.~\eqref{CondProbCC} for such an operator, then they cannot be explained by common-cause alone, indicating that there must be at least some measure of direct causal influence as well.

In the purely direct-cause scenario, the causal map is of the form
\begin{equation}
\mathcal{E}^{\rm dc}_{CB|D}= \rho_{C} \otimes \mathcal{E}_{B|D},
\end{equation}
for some quantum channel $\mathcal{E}_{B|D}$ and some state $\rho_C$.
It follows that in this case, 
\begin{equation}\label{dcstatistics}
P(km|su) = {\rm Tr}_{C} \left( \Pi_{sk}^C \rho_C \right) {\rm Tr}_{BD} \left[ \Pi_{um}^B  \mathcal{E}_{B|D}(\Pi_{sk}^D) \right].
\end{equation}

Tracing over the output of the channel, $B$, gives ${\rm Tr}_{B} \left(   \mathcal{E}_{B|D}(\Pi_{sk}^D) \right)=1$ regardless of the input $\Pi_{sk}$ on $D$, so the marginal probability distribution over $k$ becomes
\begin{equation}\label{margdistdc}
P(k|su)={\rm Tr}_{C} \left( \Pi_{sk}^C \rho_{C} \right)=P(k|s),
\end{equation}
which depends only on the marginal state $\rho_C$, as it did in the common-cause scenario, Eq.~\eqref{margdistcc}. The conditional distribution in the direct-cause case is therefore given by
\begin{equation}
P(m|ksu) = {\rm Tr}_{BD} \left[ \Pi_{um}^B  \mathcal{E}_{B|D}(\Pi_{sk}^D) \right],
\end{equation}
determined entirely by the channel $\mathcal{E}_{B|D}$.

In order to facilitate the comparison with the common-cause scenario, we express the conditional probability in terms of the Jamio\l{}kowski-isomorphic operator of the channel $\mathcal{E}_{B|D}$:
\begin{equation}\label{CondProbDC}
P(m|ksu) = {\rm Tr}_{BD} \left(  \Pi_{um}^{B} \otimes \Pi_{sk}^{D} \rho_{B|D}\right),
\end{equation}
which closely parallels the form of Eq.~\eqref{CondProbCC}.
Operators like $\rho_{B|D}$ are introduced in Ref.~\cite{LeiferSpekkens} as \emph{causal} conditional states. Like acausal conditionals above, in order to represent a trace-preserving map, they satisfy the law of total probability, 
\begin{equation}
{\rm Tr}_{B} \rho_{B|D} = \mathbb{1}_{D}.
\end{equation}
However, in order to represent a \emph{completely positive} map, $\rho_{B|D}$ must have a positive partial transpose (PPT) rather than being positive itself.  This is the key difference from the common-cause case, which ultimately allows us to distinguish the two scenarios.
Specifically, if the statistics $P(m|ksu)$ \emph{cannot} be cast in the form of Eq.~\eqref{CondProbDC} for a PPT operator $\rho_{B|D}$, then they cannot be explained by a purely direct-cause mechanism, indicating that there must be at least some measure of common cause as well.

To summarize, if the observed statistics cannot be put in the form \eqref{CondProbCC} with $\rho_{B|C}$ positive semi-definite, then they do not admit of an explanation purely in terms of a common cause, while if they cannot be put in the form of Eq.~\eqref{dcstatistics}, with $\rho_{B|D}$ PPT, then they do not admit of an explanation purely in terms of a direct cause.  Hence, properties of the observed statistics can not only witness the impossibility of a common cause explanation, as noted in Ref.~\cite{Fitzsimons2013}, they can also witness the impossibility of a direct cause explanation.  It is in this sense that the observed statistics in a passive observation scheme contain signatures of the causal structure. Note that this analysis applies regardless of the dimension of the Hilbert spaces describing the systems.

It is interesting to note that, if $C$ and $B$ are related by a common cause, then the marginal probability distributions over the outcomes of both measurements, $k$ and $m$, are independent of the setting of the respective other measurement: it follows from Eq.~\eqref{ccstatistics} that
\begin{eqnarray}
P(k|su)&=&P(k|s),\label{CI1}\\
P(m|su)&=&P(m|u).\label{CI2}
\end{eqnarray}
This captures the impossibility of signalling between two systems that are not connected by a direct-cause link.

In the direct-cause case, Eq.~\eqref{margdistdc} implies that
\begin{equation}
P(k|su)=P(k|s)\label{CI3},
\end{equation}
that is, the outcomes of measurements on $C$ are independent of the setting at $B$, which is a consequence of the fact that $C$ is prior to $B$ in time. However, the marginal distribution over the outcome at $B$,
\begin{equation}
P(m|su) = \sum_{k} {\rm Tr}_{C} \left( \Pi_{sk}^C \rho_C \right) {\rm Tr}_{BD} \left[ \Pi_{um}^B  \mathcal{E}_{B|D}(\Pi_{sk}^D) \right],
\end{equation}
is generally \emph{not} independent of the measurement setting $s$ at $C$, allowing one to signal via the direct-cause connection. 

In the context of causal inference, observing such signalling ($P(m|su) \ne P(m|u)$) allows one to trivially conclude that there must be at least some measure of direct causal influence. Conversely, if we wish to study the case wherein the quantum causal inference problem is nontrivial, we can restrict ourselves to situations that preclude signalling. Namely, if one has no prior knowledge of $C$, so that $\rho_C$ is the maximally mixed state, $\rho_C = \frac{1}{d_C} \mathbb{1}_C$,  
then $P(k|s)=\frac{1}{d_C}$ and consequently
\begin{eqnarray}
P(m|su)
&=& \frac{1}{d_C} {\rm Tr}_{B} \left[ \Pi_{um}^B  \mathcal{E}'_{B|D}(\sum_k  \Pi_{sk}^D) \right]\nonumber\\
&=& \frac{1}{d_C} {\rm Tr}_{B} \left[ \Pi_{um}^B  \mathcal{E}'_{B|D}(\mathbb{1}_D) \right]
\end{eqnarray}
which implies
\begin{equation}
P(m|su)=P(m|u)\label{CI4}.
\end{equation}
The latter condition asserts that there is no possibility of signalling by choosing the setting $s$ and observing the outcome $m$. On the other hand, if $\rho_C$ is not the maximally mixed state, different measurement settings $s$ give rise to different distributions $P(k|s)$, which ultimately allow signalling. Thus, for the purpose of causal inference, we can restrict ourselves to the case of $\rho_C$ maximally mixed, since otherwise the problem admits a trivial solution.

Furthermore, if we consider a probabilistic mixture of common-cause and direct-cause relations, 
\begin{equation}
\rho_{CB|D} = p  \rho_{CB} \otimes \mathbb{1}_D +(1-p) \rho_C \otimes \rho_{B|D},
\end{equation}
and the marginal $\rho_C$ of the direct-cause component is assumed to be maximally mixed so as to prevent signalling, then we can also assume that the marginal of the bipartite state, ${\rm Tr}_B \rho_{CB}$, is maximally mixed. 
If this were not the case, the marginal statistics $P(k|s)$ would be sufficient to distinguish the two causal scenarios. (The marginal state on $B$ in both the direct-cause and common-cause cases remains arbitrary and therefore the marginal statistics $P(m|u)$ cannot distinguish the two cases either.) 

To summarize, in the nontrivial version of the quantum causal inference problem, the conditional independences that hold in the direct-cause scenario, Eqs.~\eqref{CI3} and \eqref{CI4}, are precisely the same as those that hold in the common-cause scenario, Eqs.~\eqref{CI1} and \eqref{CI2}, and the possibilities for the marginal statistics $P(k|s)$ and $P(m|u)$ are also precisely the same in the two scenarios. The nontrivial version of the problem is the one wherein the signature of the causal structure must be found in the form of the correlations alone, that is, in the form of $P(m|ksu)$.

\subsection{Reconstruction of the causal map given a promise}\label{promise}

We now demonstrate that the passive observation scheme sometimes allows a complete solution of the quantum causal inference problem. More specifically, we show that if the systems are qubits and one is promised that the unknown circuit fragment is a probabilistic mixture of a direct cause mechanism associated with a unitary and a common-cause mechanism associated with a pure maximally entangled state, then passive observation is sufficient to solve the problem (up to a binary ambiguity in the general case). 

As discussed in the previous section, causal inference based on passive observation relies on the conditional probability distribution $P(m|ksu)$. In the case of a probabilistic mixture of common-cause and direct-cause, this can be expressed in terms of two conditional quantum states, $\rho_{B|C}$ and $\rho_{B|D}$, as
\begin{eqnarray}\label{hybridstatistics}
P(m|ksu) 
&=& p {\rm Tr}_{BC} \left( \Pi_{um}^B \otimes \Pi_{sk}^C \rho_{B|C} \right)\nonumber\\
&+& (1-p) {\rm Tr}_{BD} \left(  \Pi_{um}^{B} \otimes \Pi_{sk}^{D} \rho_{B|D}\right).
\end{eqnarray}
The promise that defines our more restricted causal inference problem is: (i) the bipartite state that gives rise to the common-cause conditional, $\rho_{BC}$, is pure and maximally entangled. (In the context of the constraint that $\rho_{BC}$ has a maximally mixed marginal on $C$, we note that $\rho_{B|C}=2\rho_{CB}$, and it is sufficient to demand that $\rho_{BC}$ be maximally entangled.) (ii) $\rho_{B|D}$ describes a unitary process.
Achieving a complete solution of the causal inference problem in this case corresponds to determining $\rho_{B|C}$, $\rho_{B|D}$ and the value of the mixing parameter $p$.  Here we derive an explicit and unique (up to one choice of sign) solution to this problem. The analysis refers only to qubits, because some steps do not apply to higher dimensions (such as the Bloch sphere representation and Euler's rotation theorem). 

\begin{figure*} 
  \centering
  \includegraphics[width=2.0\columnwidth]
{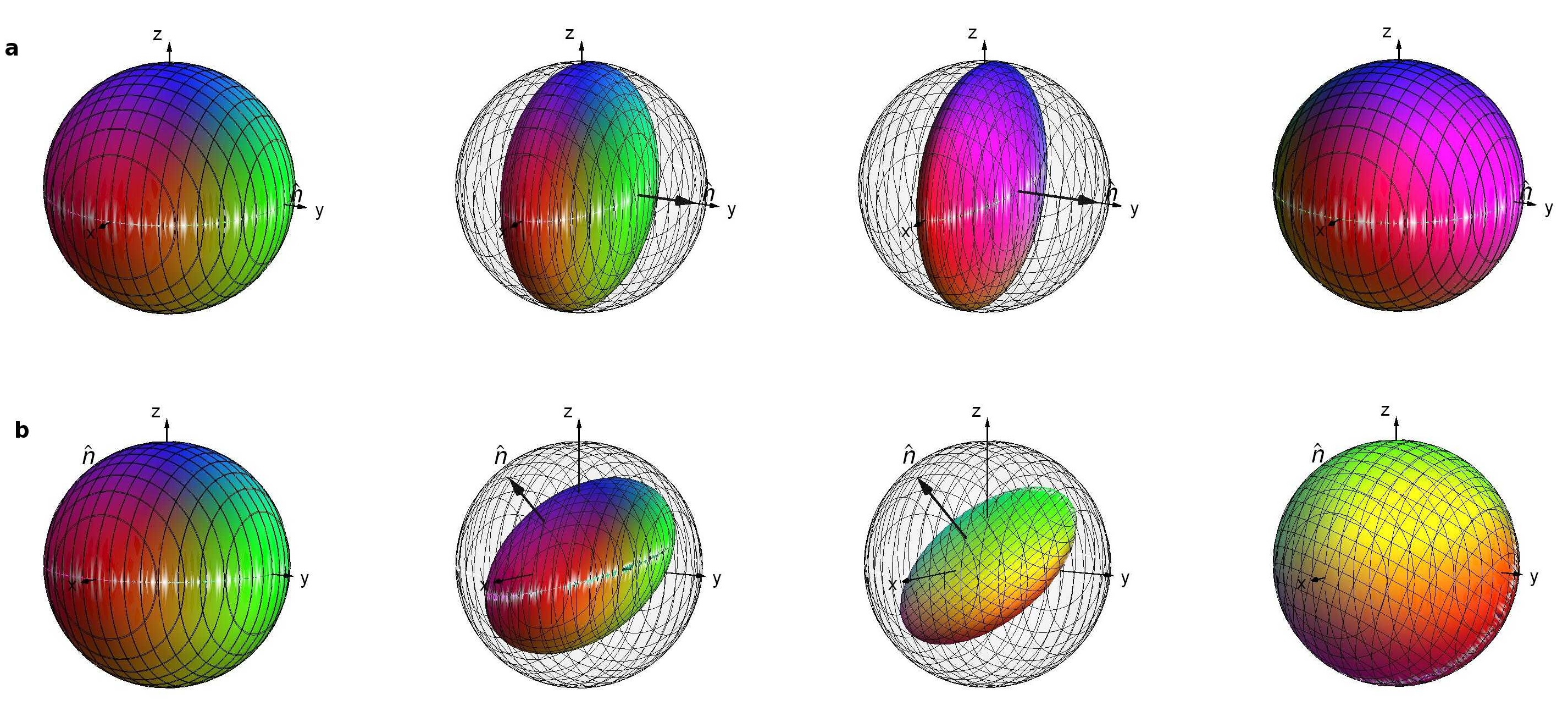} 
  \caption{\footnotesize{ 
{\bf Bloch sphere representation of a probabilistic mixture of a unitary process and a pure, maximally entangled state.} The image of the Bloch sphere under an affine map is an ellipsoid, with colours distinguishing the images of different inputs: red denotes the image of $|+_x \rangle$, cyan (anti-red) for $|-_x \rangle$, green for $|+_y \rangle$ and so on. 
Unitary processes correspond to a unit sphere with the colours distributed such that the triad red-green-blue is right-handed, such as the identity channel, on the far left, while pure, maximally entangled states correspond to unit spheres with a left-handed distribution, as they appear on the far right. Mixtures of the two extremes (shown for probability of common-cause $p=0.00, 0.25, 0.65, 1.00$, from left to right) produce ellipsoids that are flattened in the direction $\hat{n}$ (thick arrow) to a height $2\left|1-2p\right|$, and rotationally symmetric in the plane orthogonal to $\hat{n}$, with radius $r$. (a) A mixture of the identity channel and the state $|\Phi^+\rangle$, as realized in our experiment,
corresponds to the axis $\hat{n}$ pointing along $\hat{y}$, and a radius in the plane orthogonal to $\hat{n}$ of $r=1$ throughout the transition.
 (b) Mixing the identity channel with a generic pure, maximally entangled state produces intermediate ellipsoids with $\hat{n}$ pointing in a generic direction and radius $r\le1$.
} }
\label{ellipsoids}
\end{figure*}

Our proof technique is formulated in terms of a particular representation of single-qubit processes and the steering associated with two-qubit states \cite{Jevtic2013}, namely the effect that they have on the Bloch sphere. Both cases correspond to an affine transformation that rotates and scales the Bloch vectors, and, in the case of non-unital processes and non-maximally entangled states respectively, introduces an offset.  This representation can be easily visualized: for each point on the Bloch sphere, one plots the corresponding point in the Bloch sphere that is the image of the first point under the map, resulting in an ellipsoid. In order to describe the map completely, the ellipsoid can be colour-coded: the colour of a point on the ellipsoid indicates which point on the Bloch sphere it is the image of, for instance the image of $|+x\rangle$ is coloured red, the image of $|-x\rangle$ is cyan (anti-red), green for $|+y\rangle$ and so on. Fig.~\ref{ellipsoids} shows several examples of such ellipsoids.

An analytical description of the affine transformation is provided by the Pauli basis components of the Jamio\l{}kowski operator of the map.  For the direct-cause case,
\begin{equation}
\Theta_{s^{\prime}u^{\prime}}\equiv\text{Tr}\left(\rho_{B|D} \sigma_{u^{\prime}}^{B} \otimes \sigma_{s^{\prime}}^{D}\right),
\end{equation}
and for the common-cause case,
\begin{equation}
\Theta_{s^{\prime}u^{\prime}}\equiv\text{Tr}\left(\rho_{B|C} \sigma_{u^{\prime}}^{B} \otimes \sigma_{s^{\prime}}^{C}\right),
\end{equation}
where $s^{\prime},u^{\prime}\in\left\{ 0,1,2,3\right\} $ index the basis elements $\left\{ \mathbb{1},\sigma_{1},\sigma_{2},\sigma_{3}\right\}$. 

The components $\left\{ \Theta_{0,u}\right\} _{u=1,2,3}\equiv\vec{c}$ encode the offset of the centre of the ellipsoid. In the direct-cause case, this Bloch vector describes the image under the quantum channel of the maximally mixed state. In the common-cause case, it describes the state on $B$ unconditioned on any measurement outcome on $C$. For the unitary channels and maximally entangled states that we are considering, $\vec{c}$ is the zero vector.

The components with non-zero indices define the matrix 
\begin{equation}
T\equiv\left\{ \Theta_{su}\right\} _{s,u=1,2,3},
\end{equation}
which encodes the rotation and scaling of the ellipsoid. More specifically, it can be shown that the directions of the axes of the ellipsoid are given by the eigenvectors of $TT^{T}$, and their lengths by the square roots of its eigenvalues. 

In this context, we term a single-qubit channel or a two-qubit state \emph{extremal} if its ellipsoid coincides with the unit sphere.
In the case of channels, this is a well-known constraint: it implies that the channel must introduce no noise, but rather implements a pure rotation described by a unitary operator. Meanwhile, two-qubit states whose steering ellipsoid is the full sphere are pure and maximally entangled. 

In the extremal case, the distinction between states and processes is simple: extremal processes correspond to proper rotations of the Bloch sphere, with $\det T=+1$. If the colour-coding is such that the triad red-green-blue is right-handed on the input sphere, then the coloured sphere describing the output of the channel will have that same handedness. Extremal \emph{bipartite states}, on the other hand, produce steering ellipsoids whose colour distribution has the opposite handedness to the input sphere, being reached by an improper rotation, with $\det T=-1$. (This can be seen algebraically as follows: partial transposition, which maps the set of causal conditional states to the set of acausal conditional states, simply changes the sign of terms involving $\sigma_{2}$, which implies multiplying the $T$ matrix by $diag\left\{ +1,-1,+1\right\} $.) The examples discussed in Table 1 of the main article are instances of extremal states and processes.

\begin{figure}
\begin{centering}
\includegraphics[width=0.85\columnwidth]{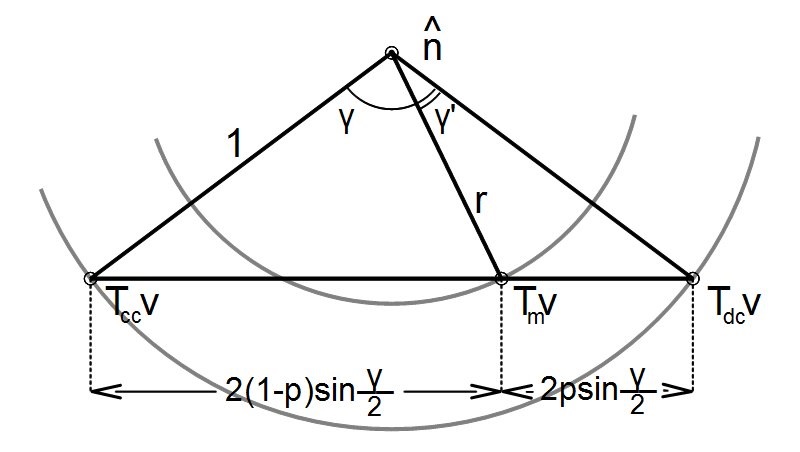}
\par
\end{centering}

\caption{ {\bf Geometric construction for characterizing a probabilistic mixture of a unitary process and a pure, maximally entangled state.}
In the plane orthogonal to $\hat{n}$, the image of a given input Bloch vector $\vec{v}$ under $T_{m}$ lies on the chord connecting the images under $T_{dc}$ and $T_{cc}$. Its distance from the centre, which gives the radius $r$, is related to the angle $\gamma$ spanned by the chord and the probability $p$ of common cause in the mixture.}
\label{geometric}
\end{figure}

Before turning to a probabilistic mixture of a generic extremal process and a generic extremal state, consider how the two are related. Let the matrices $T_{dc},T_{cc}$ encode their respective effects on the Bloch sphere, and
note that $T_{dc}$ can be transformed into $T_{cc}$ in two steps: reflection through the origin, which we denote by $F$, and rotation about some particular axis $\hat{n}$ by some particular angle. We write the angle as $\pi+\gamma$, so that the rotation can be decomposed into $R_{\hat{n},\pi}$ followed by $R_{\hat{n},\gamma}$.
Thus 
\begin{equation}
T_{cc}=R_{\hat{n},\gamma}R_{\hat{n},\pi}F T_{dc}.
\end{equation}
After the first step, the image of each point on the surface of the sphere under $F T_{dc}$ is diametrically opposed to its image under $T_{dc}$ alone. 
Under the rotations, the two points whose images lie at $\pm\hat{n}$ remain opposites. Meanwhile, the images in the plane orthogonal to $\hat{n}$ coincide again after the $\pi$ rotation.
Once we include the final rotation by $\gamma$, there will be an offset by $\gamma$ between the images of $T_{dc}$ and $T_{cc}$ in the plane orthogonal to $\hat{n}$, while their images along $\hat{n}$ are diametrically opposed. 

Now consider a probabilistic mixture of the two extremal cases.  By linearity, it is associated to a matrix
\begin{equation}
T_{m}\equiv(1-p)T_{dc}+pT_{cc}.
\end{equation}
The image of the Bloch sphere under such a combination is shown in Fig.~\ref{ellipsoids}. It must still be an ellipsoid, since $T_{m}$ is an affine transformation. Furthermore, it inherits the symmetry under rotation about $\hat{n}$. Therefore it has one semi-axis (eigenvector of $T_mT_m^{T}$) along $\hat{n}$, and a degenerate pair orthogonal to it. The length of the semi-axis (square root of eigenvalue) along $\hat{n}$ is $\left|1-2p\right|$, because the images under $T_{dc}$ and $T_{cc}$ along this direction are diametrically opposed. When $p=\frac{1}{2}$, this implies that the ellipsoid reduces to a disk. For $p<\frac{1}{2}$, the contribution from the process dominates, so $\det T_m >0$, while $\det T_m <0$ heralds $p>\frac{1}{2}$. The length of the other two semi-axes, in the plane orthogonal to $\hat{n}$, can be obtained using the geometrical construction in Fig.~\ref{geometric}:
\begin{equation}
\sin^{2}\frac{\gamma}{2}=\frac{1-r^{2}}{4\left(p-p^{2}\right)}.
\end{equation}
The images of points under $T_m$ that lie in this plane are rotated from the corresponding images under $T_{dc}$ by an angle $\gamma^\prime$, in the same direction (same sign) as $\gamma$ above, and with magnitude given by
\begin{equation}
2r\cos\gamma^{\prime}=1^{2}+r^{2}-\left[2p\sin\frac{\gamma}{2}\right]^2.
\end{equation}

Given an ellipsoid that arose from such a convex combination, it is straightforward to extract the direction $\hat{n}$, the probability $p$ and the angle $\gamma$. (In the pathological case that all three semi-axes have the same length, one finds that $\gamma=0$, which implies that the image of $T_{cc}$ is diametrically opposed to that of $T_{dc}$ for all inputs, and there is no need to single out a direction $\hat{n}$. The probability $p$ can still be read off normally.)

Given those parameters, the following steps then allow one to recover $T_{dc}$ from $T_m$: (1) scaling by $1/(1-2p)$ in the direction of $\hat{n}$ and $1/r$ in the perpendicular plane, which, as a matrix operation, we denote by $S_{\bot \hat{n},1/r} S_{\hat{n},1/(1-2p)}$; and (2) rotation about $\hat{n}$ by $-\gamma^\prime$, $R_{\hat{n},-\gamma^\prime}$.  In all, we have
\begin{equation}
T_{dc}= R_{\hat{n},-\gamma^\prime} S_{\bot \hat{n},1/r} S_{\hat{n},1/(1-2p)} T_m.
\end{equation}
Similarly, the common-cause contribution can be found via
\begin{equation}
T_{cc}= R_{\hat{n},- \gamma + \gamma^\prime } S_{\bot \hat{n},1/r} S_{\hat{n},1/(2p-1)} T_m.
\end{equation}
Note that there is an ambiguity in the direction of the rotations: assuming we take $\gamma$ and $\gamma^\prime$ to be non-negative by convention, whether $\hat{n}$ points in one or the other direction along the axis it defines (which is not fixed by the ellipsoid) generates two possible solutions. However, they are related by simple rotations of $T_{cc}$ and $T_{dc}$ about $\hat{n}$ by fixed angles.

Note also that the ambiguity is removed when $\gamma=0$, as is the case in the example that we implemented in our experiment, mixing the identity channel and the state $|\Phi^+\rangle$. But even if this ambiguity persists, given a probabilistic mixture of any unitary process and any maximally entangled pure bipartite state, we can uniquely determine the mixing probability $p$ as well as the angle $\gamma$ and the direction of $\hat{n}$ up to an inversion about the origin.


\subsection{The role of coherence and entanglement}\label{coherence}

The previous section shows that \emph{extremal} states and processes can be distinguished perfectly by passive observation alone, and if one is promised a probabilistic mixture of two such objects, the causal inference problem can be solved completely without the need for interventions. 

Without the promise of extremality, however, passive observation does not allow a complete solution of the causal inference problem. Indeed, when confronted with a channel that is entanglement-breaking, or a bipartite state that is separable, passive observation does not provide \emph{any} information about the causal relations. This is because the sets of correlations that can arise from the two scenarios are identical, corresponding to conditional states that are both positive and PPT.  In these cases, we have the same ambiguity as is found classically.  These cases can be identified by their steering ellipsoid using a criterion due to Jevtic et al. \cite{Jevtic2013}: a two-qubit state is separable and a single-qubit channel is entanglement-breaking if and only if the ellipsoid it defines 
fits inside a tetrahedron which is in turn circumscribed by the Bloch sphere.

It follows that in order to obtain some information about the causal relations, it is necessary that one or both of the following conditions hold:  the direct-cause mechanism is a channel that preserves some coherence, or the common-cause mechanism is a bipartite state that has some entanglement. We conclude that having either coherence or entanglement is a necessary conditions for achieving a quantum advantage for causal inference.

Conversely, coherence of the process and entanglement of the bipartite state is also a sufficient condition for inferring the causal structure, if we are promised that it is either purely common-cause or purely direct-cause, and the systems are qubits. To see this, recall that two-qubit states are separable if and only if they have a positive partial transpose. If the state is known to be entangled, its partial transpose must not be positive, $\rho_{B|C}^{T_C}\ngeq 0$, which implies a pattern of correlations that rules out a purely direct-cause explanation. Similarly, single-qubit processes that are not entanglement-breaking have $\rho_{B|D}\ngeq 0$, which rules out a purely common-cause explanation. Thus we can identify the causal structure unambiguously.

\section{Reconstructing the causal map from experimental data}

We have shown that for the interventionist scheme, we can reconstruct the causal map from the observed statistics, and in the passive observation scheme, we can do so if we are given a promise about the form of the causal map.  In an experiment with a finite number of runs, the observed statistics are subject to statistical fluctuations, therefore one estimates the causal map by a least-squares fit procedure.  That is, one determines the causal map that generates statistics that are closest, according to a particular figure of merit, to the observed statistics.  In this section, we describe some of the details of this fitting procedure. We first consider data from the interventionist scheme, then present the analysis of data obtained by passive observation.

While the theoretical analysis in the previous sections was based on relative frequencies of outcomes given settings, such as $P(km|lstu)$, the statistical analysis of experimental data is based directly on the numbers of counts obtained for each combination of parameters. We denote the observed absolute frequencies by $\tilde{P}^{\rm obs}(km|lstu)$. This is the number of coincidence counts detected when the wave-plates before and after the polarizers are set to implement a certain set of values of $klmstu$. 
Assuming that each set of values $klmstu$ was implemented on $N$ runs of the experiment, we can write the count numbers predicted by the fitting model as
\begin{equation}
\tilde{P}^{\rm fit}(km|lstu)= N P^{\rm fit}(km|lstu).
\end{equation}

The relative frequencies $P^{\rm fit}(km|lstu)$ are given in terms of the causal map $\rho_{CB|D}$ by
\begin{equation}
P^{\rm fit}(km|lstu)={\rm Tr}_{CBD} \left[ \rho_{CB|D}(\Pi_{sk}^{C}\otimes \Pi_{um}^{B}\otimes \Pi_{tl}^{D}) \right].
\end{equation}
We do not consider the most general form of a causal map, but use the promise that the causal structure is a probabilistic mixture of common-cause and direct-cause:
\begin{equation}
\rho_{CB|D} = p  \rho_{CB} \otimes \mathbb{1}_D +(1-p) \rho_C \otimes \rho_{B|D},
\end{equation}
where $\rho_{CB}$ and $\rho_{C}$ are density operators, ie trace-one positive-semidefinite, and $\rho_{B|D}$ is PPT and satisfies the law of total probability, since it represents a channel.
Thus, we can consider our model to be parameterized by $p$, $N$ and the operators $\rho_{CB}$, $\rho_{C}$, $\rho_{B|D}$, with the predicted count numbers given by
\begin{eqnarray}
\tilde{P}^{\rm fit}(km|lstu)=pN {\rm Tr}[\rho_{CB} ({\Pi_{sk}^C} \otimes {\Pi_{um}^B})] \nonumber\\
+(1-p)N  {\rm Tr}[\rho_{C} {\Pi_{sk}^C}] {\rm Tr}[\rho_{B|D} ({\Pi_{um}^B} \otimes {\Pi_{tl}^D})] .
\end{eqnarray}

The fitting is simplified if, rather than imposing the appropriate normalization of all the operators and subsequently including additional parameters $N$ and $p$, we allow one operator in each term to be unnormalized: we define
\begin{eqnarray} \label{defrhotilde}
\tilde{\rho}_{CB}\equiv pN \rho_{CB}\\
\tilde{\rho}_{C}\equiv (1-p)N \rho_{C},
\end{eqnarray}
in terms of which
\begin{eqnarray}
\tilde{P}^{\rm fit}(km|lstu)= {\rm Tr}[\tilde{\rho}_{CB} ({\Pi_{sk}^C} \otimes {\Pi_{um}^B})] 
\nonumber\\
+  {\rm Tr}[\tilde{\rho_{C}} {\Pi_{sk}^C}] {\rm Tr}[\rho_{B|D} ({\Pi_{um}^B} \otimes {\Pi_{tl}^D})].
\end{eqnarray}
We seek the model, parametrized by $\tilde{\rho}_{CB}$, $\tilde{\rho}_C$ and $\rho_{B|D}$, that best fits the observed frequencies, in the sense that it minimizes the residue
\begin{equation}\label{chisquared1}
\chi^2=\sum_{klmstu} \frac{\left[ \tilde{P}^{\rm fit}(km|lstu) - \tilde{P}^{\rm obs}(km|lstu) \right] ^2}{\tilde{P}^{\rm fit}(km|lstu)}.
\end{equation}

Recall from Section \ref{causaltomo} that $\rho_{CB|D}$ can be expressed as a function of the statistics one would obtain in the limit of infinitely many runs.  It follows that we expect the least-squares fit of experimental data to find a unique global minimum, with the best-fitting $\rho_{CB|D}$ close to the one realized in the experiment.

In the case of passive observation, the observed absolute frequencies are $\tilde{P}^{\rm obs}(km|su)$. The absolute frequencies predicted by the model are related to the relative frequencies by the number of runs for each set of values of the settings $kmsu$, which we denote $N$:
\begin{equation}
\tilde{P}^{\rm fit}(km|su)= N P^{\rm fit}(km|su).
\end{equation}
Considering a probabilistic mixture of common-cause and direct-cause, and recalling that, in the passive observation scheme, we reprepare the same state $\Pi_{sk}$ on $D$ that was found on $C$, the relative frequencies are
\begin{eqnarray}
P^{\rm fit}(km|su)=p {\rm Tr}[\rho_{CB} ({\Pi_{sk}^C} \otimes {\Pi_{um}^B})] \nonumber \\
+(1-p)  {\rm Tr}[\rho_{C} {\Pi_{sk}^C}] {\rm Tr}[\rho_{B|D} ({\Pi_{um}^B} \otimes {\Pi_{sk}^D})],
\end{eqnarray}
where again $\rho_{CB}$ and $\rho_C$ are states, while $\rho_{B|D}$ represents a channel.
Combining the two previous equations and using the unnormalized operators defined in Eq.~\eqref{defrhotilde},
we can write the count numbers predicted by the model for the case of passive observation as
\begin{eqnarray}
\tilde{P}^{\rm fit}(km|su)= {\rm Tr}[\tilde{\rho}_{CB} ({\Pi_{sk}^C} \otimes {\Pi_{um}^B})] \nonumber \\ 
+  {\rm Tr}[\tilde{\rho_{C}} {\Pi_{sk}^C}] {\rm Tr}[\rho_{B|D} ({\Pi_{um}^B} \otimes {\Pi_{sk}^D})].
\end{eqnarray}
We seek to minimize the residue
\begin{equation}
\chi^2=\sum_{kmsu} \frac{\left[ \tilde{P}^{\rm fit}(km|su) - \tilde{P}^{\rm obs}(km|su) \right] ^2}{\tilde{P}^{\rm fit}(km|su)}.
\end{equation}

As we have shown in Section \ref{promise}, if there is a promise that $\rho_{CB|D}$ represents a probabilistic mixture of a pure maximally entangled bipartite state and a unitary channel,
then $\rho_{BC|D}$ can be obtained from the data in the passive observation scheme up to a sign ambiguity. It follows that for our experiment, which aims to prepare a mixture of this type which removes the ambiguity, we expect there to be a unique global minimum of $\chi^2$ in parameter space, and that consequently the $\rho_{CB|D}$ which best fits the data will be close to the one realized in the experiment. Note that we do not impose maximal entanglement of $\rho_{CB}$ or unitarity of the map associated to $\rho_{B|D}$ as a constraint in our fit.  Nonetheless, as long as the experiment has come close to achieving these ideals, we expect the fit to be good over only a small interval of possibilities for $\rho_{CB}$, $\rho_{C}$, $\rho_{B|D}$, and the value of $p$.

The operators that parametrize our model are subject to the following constraints, based on properties derived in section~\ref{signatures}.
The unnormalized states $\tilde{\rho}_{CB}$ and $\tilde{\rho}_{C}$ are positive-semidefinite, and their traces are related by
\begin{equation} \label{constraintP}
 \frac{{\rm Tr}_{CB} \tilde{\rho}_{CB}}{{\rm Tr}_{CB} \tilde{\rho}_{CB}+{\rm Tr}_{C} \tilde{\rho}_{C}}=p.
\end{equation} 
The conditional $\rho_{B|D}$ is PPT and satisfies the law of total probability,
\begin{equation}\label{constraintTotalProb}
{\rm Tr}_{B} \rho_{B|D} = \mathbb{1}_D.
\end{equation}

The positive-semidefinite operators invoked above can be parametrized conveniently following Ref.~\cite{James2001}. A two-qubit positive-semidefinite operator such as $\tilde{\rho}_{CB}$ requires 16 real numbers, which we arrange into a vector $\vec{r}_{CB}$.
Define the lower-triangular 4 by 4 matrix
\begin{equation}\label{lowertriangular}
R_{CB}=\left(\begin{array}{cccc}
r_{1}&0&0&0\\
r_{5}+\imath r_{6} & r_{2}&0&0\\
r_{11}+\imath r_{12} & r_{7}+\imath r_{8} & r_{3}&0\\
r_{15}+\imath r_{16} & r_{13}+\imath r_{14} & r_{9}+\imath r_{10} & r_{4}
\end{array}\right),
\end{equation}
and take
\begin{equation}
\tilde{\rho}_{CB}=R_{CB}^{\dagger}R_{CB}.
\end{equation}
This form, known as the Cholesky decomposition, is manifestly positive-semidefinite, and by varying over the vectors $\vec{r}_{CB}$, we vary over all $\tilde{\rho}_{CB}$. A further convenient feature of this parametrization is that the trace of $\rho_{CB}$ is given simply by square of the 2-norm of $\vec{r}_{CB}$,
\begin{equation}
{\rm Tr}_{CB} \tilde{\rho}_{CB}=|\vec{r}_{CB}|^2.
\end{equation}

Positive-semidefinite operators for a single qubit can be obtained by a similar construction: a vector of 4 real numbers, $\vec{r}_C$, defines the components of a lower triangular 2 by 2 matrix $R_{C}$, in terms of which $\rho_C =  R_{C}^{\dag} R_{C}$. 

If a PPT operator $\rho_{B|D}$ is called for, one can take the partial transpose of the above form: $\vec{r}_{B|D}$ is again a vector of 16 real numbers that defines a matrix $R_{B|D}$ by Eq.~\eqref{lowertriangular}, and one takes $\rho_{B|D} = (R_{B|D}^{\dagger}R_{B|D})^{T_D}$, where $T_D$ is the partial transpose. We note that the trace of the PPT operator is also given by the square of the norm of its parameter vector: ${\rm Tr}_{BD} \tilde{\rho}_{B|D}=|\vec{r}_{B|D}|^2.$

The additional constraints on operators are enforced by adding penalty functions to the principal function $\chi^2$. The constraint that relates the traces of the unnormalized operators in each term to the probability $p$, Eq.~\eqref{constraintP}, can be cast directly in terms of the norms of the parameter vectors: we add a term
\begin{equation}
\lambda \left( \frac{|\vec{r}_{CB}|^2}{|\vec{r}_{CB}|^2+|\vec{r}_{C}|^2}-p\right)^2.
\end{equation}
The additive term enforcing the law of total probability, Eq.~\eqref{constraintTotalProb}, is proportional to the sum of the absolute value squared of the elements of the difference between ${\rm Tr}_B \rho_{B|D}$ and $\mathbb{1}_D$,
\begin{equation}
\lambda \sum_{ij} \left| \left( {\rm Tr}_{B} \rho_{B|D} - \mathbb{1}_D \right) _{ij} \right| ^2.
\end{equation}
The Lagrange multiplier $\lambda$ for each penalty term was selected heuristically, with values of $~10^7$ found to enforce the constraints without obscuring the principal function.

\end{document}